\documentclass[final,5p,times,twocolumn]{elsarticle}

\usepackage{graphicx}
\usepackage{dcolumn}
\usepackage{bm}
\usepackage{longtable}
\usepackage{lscape}
\usepackage{multirow}
\usepackage{booktabs}
\usepackage{amsmath}
\usepackage{float}
\usepackage{rotating}
\usepackage{caption}
\usepackage{placeins}
\usepackage{multicol}
\usepackage{xcolor}
\usepackage[colorlinks=true,linkcolor=blue,citecolor=blue,urlcolor=blue]{hyperref}
\usepackage{miller}

\journal{Acta Materialia}

\begin{document}

\begin{frontmatter}

\title{Chemically tailored planar defect phases in the Ta-Fe $\mu$-phase}

\affiliation[label1]{organization={Institute for Physical Metallurgy and Materials Physics, RWTH Aachen University, 52074 Aachen, Germany}}

\affiliation[label2]{organization={Max Planck Institute for Sustainable Materials, 40237 Düsseldorf, Germany}}

\affiliation[label3]{organization={Aachen Institute for Advanced Study in Computational Engineering Science, RWTH Aachen University, 52062 Aachen, Germany}}

\author[label1]{Christina Gasper\corref{cor1}} 
\author[label1]{Nisa Ulumuddin} 
\author[label2]{Siyuan Zhang\corref{cor1}} 
\author[label1]{Sang-Hyeok Lee} 
\author[label2]{Christina Scheu} 
\author[label3]{Benjamin Berkels} 
\author[label1]{Zhuocheng Xie\corref{cor1}} 
\author[label1]{Sandra Korte-Kerzel\corref{cor1}} 

\cortext[cor1]{Corresponding author\\
Email address: gasper@imm.rwth-aachen.de\\ 
siyuan.zhang@mpie.de\\ 
xie@imm.rwth-aachen.de\\ 
korte-kerzel@imm.rwth-aachen.de}
\begin{abstract}
Intermetallics often exhibit complex crystal structures, which give rise to intricate defect structures that critically influence their mechanical and functional properties. Despite studies on individual defect types, a comprehensive understanding of the defect landscape in $\mu$-phases, a class of topologically close-packed phases, remains elusive. In this study, we investigated the planar defect structures in the Ta-Fe $\mu$-phase across a compositional range of 46 to 58 at.\% Ta using electron microscopy and density functional theory calculations. Electron backscatter diffraction and high-resolution scanning transmission electron microscopy reveal a transition from basal twin boundaries and planar faults containing C14 TaFe$_2$ Laves phase layers at a low Ta content to pyramidal $\{1\bar{1}02\}$ twins at a higher Ta content. Density functional theory calculations of defect formation energies confirm a chemical potential-driven stabilisation of Laves phase lamellae. The prevalence of pyramidal twins in Ta-rich $\mu$-phase samples is attributed to the competitive nature of different planar defects during solidification. A defect landscape for $\mu$-phases is proposed, illustrating the interplay between site occupancy, dislocation types and planar faults across the chemical potential space. These findings provide fundamental insights into defect engineering in structurally complex intermetallics and open pathways for optimising material properties through chemical tuning.
\end{abstract}


\end{frontmatter}

\section{Introduction}
\label{intro}

Topologically close-packed (TCP) phases are a class of intermetallic compounds characterised by their complex crystal structures, which consist of two or more metallic elements with different atomic sizes \cite{sauthoff1995intermetallics}. In addition to the atomic size ratio, factors such as chemical bonding, electronegativity differences and valence electron concentration also influence the crystal structure \cite{ferro2011intermetallic, pearson1972crystal}. Due to their complexity, TCP phases tend to be brittle at room temperature \cite{sauthoff1995intermetallics, tammann1923metallographische}. However, they possess high melting points and exhibit significant strength, along with good creep and oxidation resistance at elevated temperatures, making them ideal candidates for high-temperature applications \cite{fleischer1989intermetallic, westbrook2000basic, Belin2010mechanical}. 

The $\mu$-phase, with its stoichiometric A$_6$B$_7$ composition (where A represents the larger atoms and B the smaller ones), is one of the TCP phases \cite{magneli1938rontgenuntersuchung, forsyth1962structure, kumar1998sublattice}. Its crystal structure can be described as an alternating stacking of the MgZn$_2$ Laves structure and the Zr$_4$Al$_3$ structure along the $c$-axis \cite{frank1959complex, wilson1960crystal, andersson1978structures}. Like other TCP phases, the $\mu$-phase often exhibits a broad homogeneity range that also includes the A$_7$B$_6$ composition \cite{sinha1972topologically, joubert2004mixed, luo2023tailoring}. With a large $c/a$ ratio ranging from 5.3 to 5.5 \cite{cieslak2014structural}, 39 atoms per structural unit \cite{cieslak2014structural} and a high packing density due to the presence of only tetrahedral interstices (TCP structure) \cite{shoemaker1971tetraedrisch}, the $\mu$-phase is known for its pronounced hardness and brittleness \cite{sauthoff1995intermetallics, tammann1923metallographische}. However, recent studies have explored the potential to tailor plasticity by strategically modifying local chemical distributions in the $\mu$-phase \cite{luo2023tailoring, luo2023plasticity} and by understanding how stacking and bonding within the building blocks of structurally complex crystals influence plastic deformation \cite{stollenwerk2025beyond}.

In reality, the structure of a crystal is never completely free of defects \cite{gottstein2004physical}. These defects influence the material properties and can be categorised based on their dimensionality: point defects (0D), line defects (1D), planar defects (2D), and volumetric defects (3D) \cite{gottstein2004physical}. For the $\mu$-phase,  Sluiter et al. \cite{sluiter2003site} studied the site occupancy preference in Nb-Ni $\mu$-phases using density functional theory (DFT), particularly for the 12-fold coordinated $3a$ and $18h$ sites. Their study revealed a strong Ni preference for the Kagomé layer ($18h$ sites) and a mixed occupancy in the triple layer ($3a$ sites). Building on this, Luo et al. experimentally confirmed site occupancy preferences in Nb-Co $\mu$-phase via atomic-resolution energy dispersive X-ray spectroscopy (EDS) by modifying the A:B atomic ratio \cite{luo2023tailoring}. Additionally, they explored the effects of site occupancy on basal slip mechanisms, identifying a transition from crystallographic slip via full dislocations to a glide-and-shuffle mechanism via partial dislocations as the Nb content increased from 49.5 at.\% to 54 at.\%, giving a full occupation of the $3a$ sites \cite{luo2023tailoring, luo2023plasticity}. Furthermore, Luo and Gasper et al. investigated non-basal slip in Nb-Co and Ta-Fe $\mu$-phases \cite{luo2024non}, attributing the limited non-basal plasticity to the widespread presence of basal stacking faults and the formation of cracks at intersections between non-basal and basal defects. Schr{\"o}ders et al. reported basal and prismatic slip events in a Mo-Fe $\mu$-phase \cite{schroders2018room}, where they also identified grown-in planar defects, such as basal twin boundaries arising from the absence of the Zr$_4$Al$_3$ layer \cite{schroders2019structure}. Chen et al. reported various basal planar faults in a Ta-Fe $\mu$-phase, including stacking faults formed by incorporating the MgZn$_2$ layers, intergrowth faults formed by removing or inserting a Zr$_4$Al$_3$ building block, and twin boundaries formed within the Zr$_4$Al$_3$ layer \cite{chen2025defect}. Similar basal planar faults, including stacking faults and twin boundaries, have been reported in other $\mu$-phases \cite{gao2016situ, ma2018atomic, cheng2021atomic, jin2022experimental}. Basal $(0001)$ twins predominantly form within the Zr$_4$Al$_3$ layer as confirmed by both experimental observation and DFT calculations \cite{jin2022experimental}, which indicate that these twin boundaries are energetically favourable. Additionally, pyramidal $\{1\bar{1}02\}$ and $\{\bar{1}101\}$ planar faults associated with the intergrowth of basal faults in the Ta-Fe $\mu$-phase \cite{chen2025defect}, two pyramidal $\{1\bar{1}02\}$ planar faults with distinct atomic and chemical arrangements in Co-base $\mu$-phases \cite{chen2025mechanisms}, and pyramidal $\{1\bar{1}02\}$ twin boundaries in Ni-base $\mu$-phases \cite{zhao2023atomic, gao2016situ} have been reported. However, despite these studies, a comprehensive understanding of the defect landscape in $\mu$-phases remains elusive, particularly with respect to the dependence of their formation on the chemical potential.

In this study, we investigate planar defects, with a particular focus on twin boundaries, in the binary Ta-Fe $\mu$-phase across a wide compositional range of 46 to 58 at.\% Ta. The Ta-Fe system is very well suited for this investigation, as it contains not only a $\mu$-phase with a broad homogeneity range of 14.5 at.\% but also has a C14 Laves phase, which forms one of the two building blocks of the complex $\mu$-phase crystal structure \cite{okamoto2013fe, witusiewicz2011experimental}. Electron backscatter diffraction (EBSD) and high-resolution scanning transmission electron microscopy (HR-STEM) were used to characterise the planar defects across different compositions of the Ta-Fe µ-phase. To rationalise the prevalence of these defect structures, their formation energies were calculated by DFT. By integrating our findings with previous studies on the Ta-Fe system \cite{luo2024non, gasper2025mechanical} and other $\mu$-phase systems, we propose a comprehensive defect landscape for $\mu$-phases, advancing defect engineering strategies for structurally complex intermetallics as high-performance materials.

\section{Methods}
\subsection{Sample preparation and electron microscopy analysis}

Four $\mu$-phase samples of the binary Ta-Fe system, with different target compositions of 46, 50, 54 and 58 at.\% Ta, respectively, were prepared using a small-scale laboratory arc melter (Compact Arc Melter MAM-1 from Edmund Bühler GmbH) under a protective Ar atmosphere. High-purity Ta ($>$ 99.99 \%) and Fe (99.99 \%) were used as input materials. Without additional heat treatment, the samples were ground down to a thickness of approximately 2 mm using SiC abrasive paper. The metallographic preparation was continued by mechanical grinding and polishing. For polishing, cloths with diamond particles were used first, followed by an oxide polishing suspension (OP-U). Detailed information on the synthesis, preparation and characterisation of the topologically close-packed phases in the Ta-Fe(-Al) system can be found in our previous work \cite{gasper2024preparation}.

\begin{figure*}[ht!]
    \centering
    \includegraphics[width=1\linewidth]{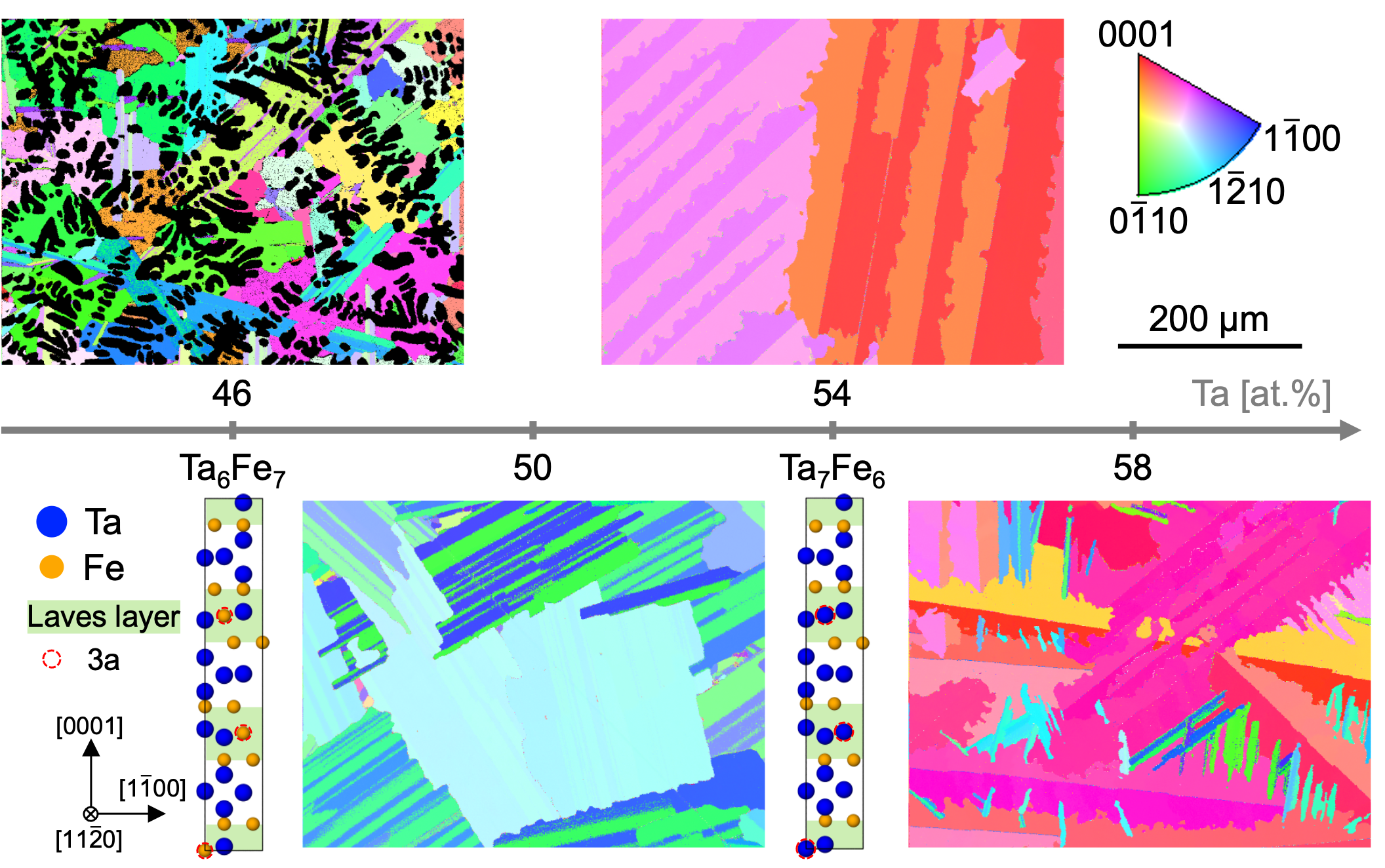}
    \caption{EBSD IPF maps of four different $\mu$-phase samples with target compositions of 46, 50, 54 and 58 at.\% Ta, rest Fe. The unindexed (black) second phase in the 46 at.\% Ta sample, identified as Laves phase, is not included in this figure. The atomic configurations of Ta$_6$Fe$_7$ and Ta$_7$Fe$_6$ $\mu$-phase unit cells are also shown, where the Laves layers are highlighted in light green and the $3a$ sites in Ta$_6$Fe$_7$ are circled in red.}
    \label{fig:1}
\end{figure*}

The prepared samples were characterised using a scanning electron microscope (SEM) equipped with EDS and EBSD detectors (Helios NanoLab 600i from FEI Inc. equipped with Octane Super A EDS detector and Hikari XP2 EBSD detector, both EDAX; EDAX TEAM software used for EDS quantification and EBSD indexation). Combined EDS and EBSD measurements were conducted at a pre-tilted sample holder of 70° with an acceleration voltage of 20 kV and a beam current of 2.7 nA. The working distance varied between 10 and 11.5 mm and a step size of 0.5 to 0.75 $\mu$m was chosen. The grain orientations of the inverse pole figure (IPF) are referenced to the sample normal. Further details on the measurement evaluation can also be found in our previous work \cite{gasper2024preparation}.

For subsequent investigation of the twin structures observed by EBSD using a HR-STEM, STEM lamellae were prepared perpendicular to the twin boundaries from the 50 and 54 at.\% Ta $\mu$-phase samples using focused ion beam (FIB) milling with a Ga ion source (Scios 2 from Thermo Fisher Scientific). First, Pt protection layers were deposited to prevent beam damage at the regions of interest, with dimensions of 10 $\mu$m in width, 1.2 $\mu$m in depth and approximately 3 $\mu$m in height. For the coarse milling of the lamellae, an acceleration voltage of 30 kV was used, while final thinning was performed using 5 kV. The final thicknesses were around 110 nm for the 50 at.\% Ta $\mu$-phase and approximately 90 nm for the 54 at.\% Ta $\mu$-phase sample. Due to the high atomic number of Ta (Z = 73) \cite{greenwood2010chemistry}, milling of the Pt protection layer was much more preferred. As a result, a lower lamella thickness could not be achieved before the protection layer was fully removed.

Both Ta-Fe $\mu$-phase twin lamellae were investigated using STEM and EDS (probe-corrected Titan Themis S/TEM  equipped with SuperX detector from Thermo Fisher Scientific), operated at 300 kV. High-angle annular dark-field (HAADF) and annular bright-field (ABF) images were recorded using detectors with angular ranges of 73 to 200 mrad and 8 to 16 mrad, respectively. Multivariate statistical analysis was performed to denoise the spectrum imaging dataset and highlight the elemental segregation \cite{zhang2018evaluation}. Each HAADF and ABF image is the average of 10 to 20 raw frames that have been aligned using non-rigid registration with bias correction \cite{BeLi19}. Here, the alignment was computed using the HAADF frames and also applied to the ABF frames.

\subsection{DFT setup and model construction}
All electronic structure calculations were performed using DFT as implemented by the Vienna $ab-initio$ Simulation Package (VASP) \cite{kresse1996efficient,kresse1993ab}. Planewave basis sets formulated by the Projector Augmented Wave (PAW) potentials were used to describe the one-electron wavefunctions, which make up the electron density of the system \cite{kresse1999ultrasoft}. These planewaves have a kinetic energy cutoff of 550 eV, and the basis sets treat Fe $4s^1, 3d^7$ and Ta $6s^1, 5d^4, 5p^6$ as the valence electrons contributing to bonding. The Perdew–Burke–Ernzerhof (PBE) exchange-correlation functional within the generalised gradient approximation (GGA) level of theory was applied \cite{perdew1996generalized}. During each geometry optimisation, the 1$^\mathrm{st}$ order Methfessel-Paxton method was implemented to smear the Fermi level with a $\sigma$ level of 0.1 eV \cite{methfessel1989high}. 
The 1$^\mathrm{st}$ Brillouin zone was sampled using the Monkhorst-Pack k-point scheme \cite{monkhorst1976special}. The self-consistent field (SCF) and geometric convergence criteria were set to 10$^{-6}$ eV and 0.02 eV/${\AA{}}$, respectively. All calculations were performed with spin polarisation. To simplify the phase space of the $\mu$-phase compositions, which emerges from the paramagnetism of the Ta-Fe material system, all models in this work were constrained to the ferromagnetic state. Based on our previous work \cite{gasper2025mechanical} and as provided in Figure S1 (in the Supplementary Material), the ferromagnetic state corresponds to either the lowest or second-lowest energy configuration for both Ta$_6$Fe$_7$ and Ta$_7$Fe$_6$.

The atomic structures of the Ta$_6$Fe$_7$ and Ta$_7$Fe$_6$ $\mu$-phases were generated using Atomsk \cite{hirel2015atomsk}. The conventional unit cells of Ta$_6$Fe$_7$ and Ta$_7$Fe$_6$ (see Figure S2 for the schematic) were optimised with an especially high k-point mesh of 10 $\times$ 10 $\times$ 5. These $\mu$-phase conventional unit cells possessing rhombohedral symmetry (R$\bar{3}$mH), have lattice parameters $a$ and $c$ with fixed angles $\alpha$ = 0$^\circ$, $\beta$ = 90$^\circ$ and $\gamma$ = 120$^\circ$. The lattice parameters were determined through systematic scans along the $a$- and $c$-directions while allowing atomic positions to relax. The lowest-energy interpolated structure was then further optimised by allowing full relaxation of the cell size, cell shape and atomic positions. The resulting DFT-optimised lattice parameters for Ta$_6$Fe$_7$ were $a$ = 4.855 $\AA{}$ and $c$ = 26.831 $\AA{}$, while for Ta$_7$Fe$_6$, they were $a$ = 4.939 $\AA{}$ and $c$ = 27.097 $\AA{}$. The calculated lattice parameters of Ta$_7$Fe$_6$ deviate by only 0.22 \% and 0.10 \% from the experimentally reported $a$ and $c$ values at room temperature, respectively \cite{raman1966mu}. 

Twin boundary and Laves planar fault models were constructed based on defect structures identified through HR-STEM. Three basal twin boundary structures were modeled: twin boundary along the Zr$_4$Al$_3$ (CN15) layer (TB$_{\mathrm{CN15}}$), twin boundary along the Kagomé layer mirroring the Laves phase layer (TB$_\mathrm{K}$(2 Laves)) and twin boundary along the Kagomé layer mirroring the Zr$_4$Al$_3$ layer (TB$_\mathrm{K}$(2 Zr$_4$Al$_3$)). Laves phase layers in the basal plane, referred to as Laves planar fault structures, were modeled by incorporating Laves phase crystal building blocks within the $\mu$-phase compositions, while maintaining the in-plane lattice of the host structures. Structures containing three and four Laves phase layers were denoted PF(3 Laves) and PF(4 Laves), respectively. The compositions of the triple layers within planar defect structures were constrained to match those of the corresponding host phase: Ta$_6$Fe$_7$ contained Ta-Fe-Ta layers, while Ta$_7$Fe$_6$ contained Ta-Ta-Ta layers. Schematics of the full simulation cells are provided in Figure S2. OVITO \cite{stukowski2009visualization} was used for visualisation and structural analysis.

\begin{figure*}[hbt!]
    \centering
    \includegraphics[width=15.2cm]{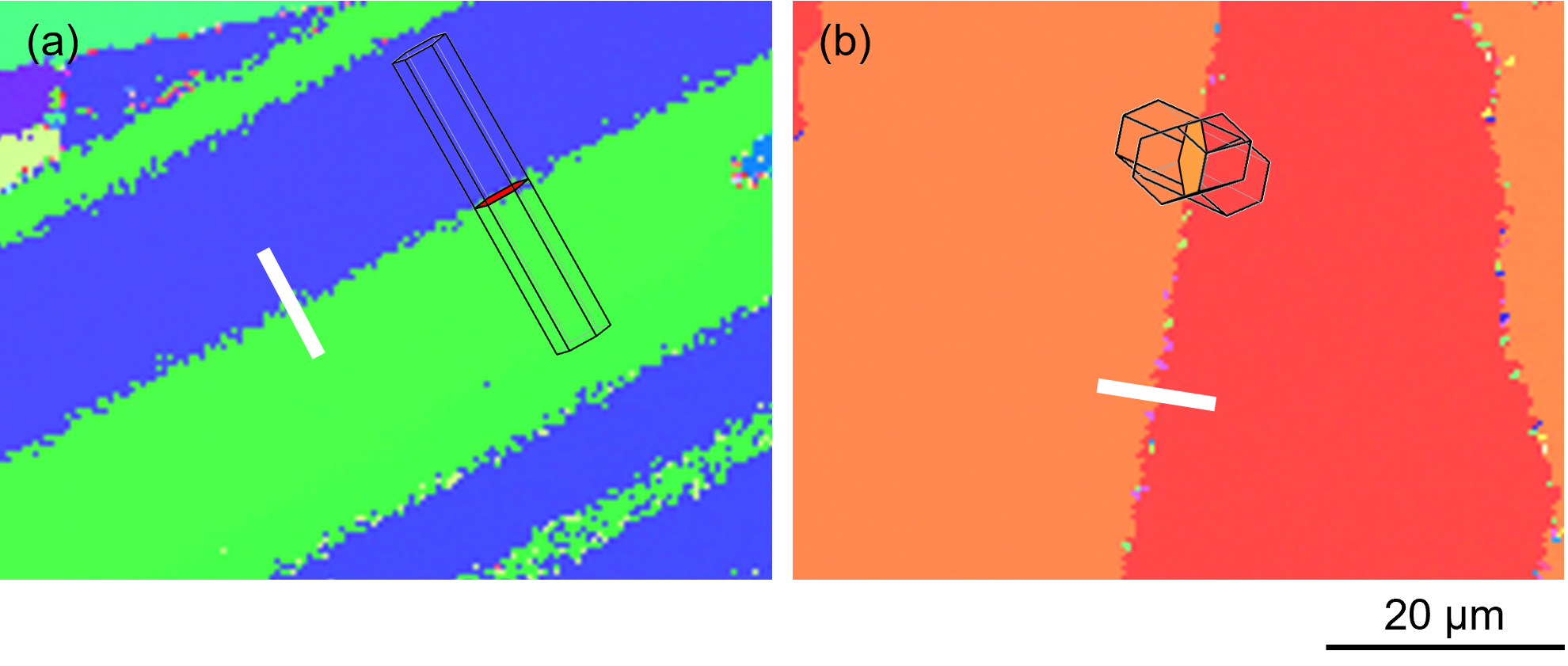}
    \caption{EBSD maps where the region for the STEM lamellae positions are highlighted in white in (a) for a basal twin in the $\mu$-phase sample with 50 at.\% Ta and in (b) for the pyramidal $\{1\bar{1}02\}$ twin in the $\mu$-phase sample with 54 at.\% Ta. The identified basal and pyramidal twin planes are highlighted in red and orange, respectively, within the overlaid oriented unit cells of the two adjacent grains.}
    \label{fig:2}
\end{figure*}

The basal twin boundary and planar fault structures were constructed with periodic boundary conditions in all directions, indicating each simulation cell contained two identical planar defects. To minimise interactions between periodic defects, each structure incorporated a 1 $\times$ 1 $\times$ 4 supercell of the pristine $\mu$-phase. Planar defects were then inserted into the host structures, extending the simulation cell along the $c$-axis. During geometry relaxation, the simulation cell was constrained to relax only along the $c$-direction (perpendicular to the planar defect), while atomic positions were fully relaxed. Consequently, the planar faults were constrained to the $a$ and $b$ lattice of the host $\mu$-phase structures. A $c$-axis scan was performed for the basal planar defect structures by incrementally varying the $c$-axis length and computing the corresponding total energies. Each data point in the scan represented a strain state along the $c$-axis. The energy corresponding to the minimised stress condition was selected as the total energy (see Figure S3 and S4). To verify that the 1 $\times$ 1 $\times$ 4 supercell is sufficiently large to avoid size-dependent artifacts arising from interactions between periodic planar faults, the defect formation energy was calculated for one representative defect structure (TB$_\mathrm{CN15}$ in Ta$_7$Fe$_6$) using a larger 1 $\times$ 1 $\times$ 6 supercell, which increases the separation between faults (see Figure S5). As the difference in defect formation energies was within the DFT resolution ($\sim$ 0.16 mJ/m$^2$),  the 1 $\times$ 1 $\times$ 4 supercell was deemed adequate. All structures were simulated with a k-point mesh of 5 $\times$ 5 $\times$ 1. The k-point benchmarking information is provided in Figure S6.


\subsection{Defect phase diagram}
To determine the energetically favourable planar defects as a function of the chemical potential, the concept of a metastable defect phase diagram was used \cite{korte2022defect}. The defect formation energy was calculated relative to the pristine $\mu$-phase, treating defect structures as localised defect regions within a thermodynamically open system with the reservoirs of its elemental components. This approach is consistent with previous studies on planar defects in the Laves phase \cite{tehranchi2024metastable}. The defect formation energy is calculated following Equation \ref{eq:defect}:

\begin{equation}
    \Delta E^\text{defect}_F = E^0_\text{defect} - E^0_\text{pristine}  - \sum_\text{i}x_\text{i}\mu_\text{i},
    \label{eq:defect}
\end{equation}

where $E^0_\text{defect}$ and $E^0_\text{pristine}$ are the total energies of the defect and the pristine $\mu$-phases, respectively. To compare the relative stability of equivalent defect structures in Ta$_6$Fe$_7$ and Ta$_7$Fe$_6$, which are treated as distinct thermodynamic phases in the Ta-Fe system, planar defects were calculated relative to the pristine $\mu$-phase that matches the composition of their host matrix. Here, $\mu_\text{i}$ and $x_\text{i}$ define the chemical potential and the number of excess solute atoms, respectively, where the solute can be Ta or Fe. 

The chemical potentials $\mu_\text{Ta}$ and $\mu_\text{Fe}$ are constrained by the stability range of the host phases, as defined by the Ta-Fe phase diagram \cite{okamoto2013fe} and experimental observations in Figure \ref{fig:1}. For the Ta$_6$Fe$_7$ phase coexisting with the C14 TaFe$_2$ Laves phase, at the lower Ta limit, the chemical potential is determined by its equilibrium with bulk Fe. At the upper Ta limit of the Ta$_6$Fe$_7$ phase, Ta$_6$Fe$_7$ coexists in equilibrium with Ta$_7$Fe$_6$. 
For the Ta$_7$Fe$_6$ phase, the lower Ta limit is defined by its equilibrium with Ta$_6$Fe$_7$, while the upper Ta limit is determined by its equilibrium with bulk Ta. 
Bulk energies are denoted by $E^0$. All chemical potentials in this work are defined under ground-state conditions ($T$ = 0 K, $P$ = 0 atm). In the defect phase diagram in this work, the Ta chemical potential axis is expressed relative to a reference state ($\Delta \mu_\text{Ta}$), which was taken as the bulk energy of body-centred cubic (bcc) Ta ($\Delta \mu_\text{Ta}$ = $\mu_\text{Ta} - \mu^{0}_\text{Ta}$). The chemical potential range for Ta in Ta$_6$Fe$_7$ and TaFe$_2$ is therefore $-0.62<\Delta \mu_\text{Ta} <-0.06$ and in Ta$_7$Fe$_6$ is $-0.06<\Delta \mu_\text{Ta} < 0.0 $.
The thermodynamic constraints of host phases are summarised in Table \ref{tab:boundaries}. 

\begin{table}[]
\caption{Thermodynamic constraints of host phases based on the Ta-Fe phase diagram and experimental observations. Bulk energies are denoted by $E^0$.   $\mu_\text{Ta}$ and $\mu_\text{Fe}$ lie within the range between these limits.}
\resizebox{\columnwidth}{!}{
\begin{tabular}{|l|c|c|}
\hline
\textbf{Host Phases}   & \textbf{\begin{tabular}[c]{@{}l@{}}Ta-poor Limit\end{tabular}} & \textbf{\begin{tabular}[c]{@{}l@{}}Ta-rich Limit\end{tabular}} \\ \hline
\textbf{TaFe$_2$ and Ta$_6$Fe$_7$} & $\mu_\text{Ta}+2\mu_\text{Fe} = E^0_\text{TaFe$_2$}$    & $6\mu_\text{Ta}+7\mu_\text{Fe} = E^0_\text{Ta$_6$Fe$_7$}$                                                                \\ 
& $\mu_\text{Fe} < E^0_\text{Fe}$ & 7$\mu_\text{Ta}+6\mu_\text{Fe} < E_\text{Ta$_7$Fe$_6$}$ \\ \hline
\textbf{Ta$_7$Fe$_6$} & 7$\mu_\text{Ta}+6\mu_\text{Fe} = E^0_\text{Ta$_7$Fe$_6$}$     &  7$\mu_\text{Ta}+6\mu_\text{Fe} = E^0_\text{Ta$_7$Fe$_6$}$ \\
& $6\mu_\text{Ta}+7\mu_\text{Fe} < E_\text{Ta$_6$Fe$_7$}$&  $\mu_\text{Ta} < E^0_\text{Ta}$\\ \hline
\end{tabular}}
\label{tab:boundaries}
\end{table}

\begin{figure*}[hbt!]
    \centering
    \includegraphics[width=0.75\linewidth]{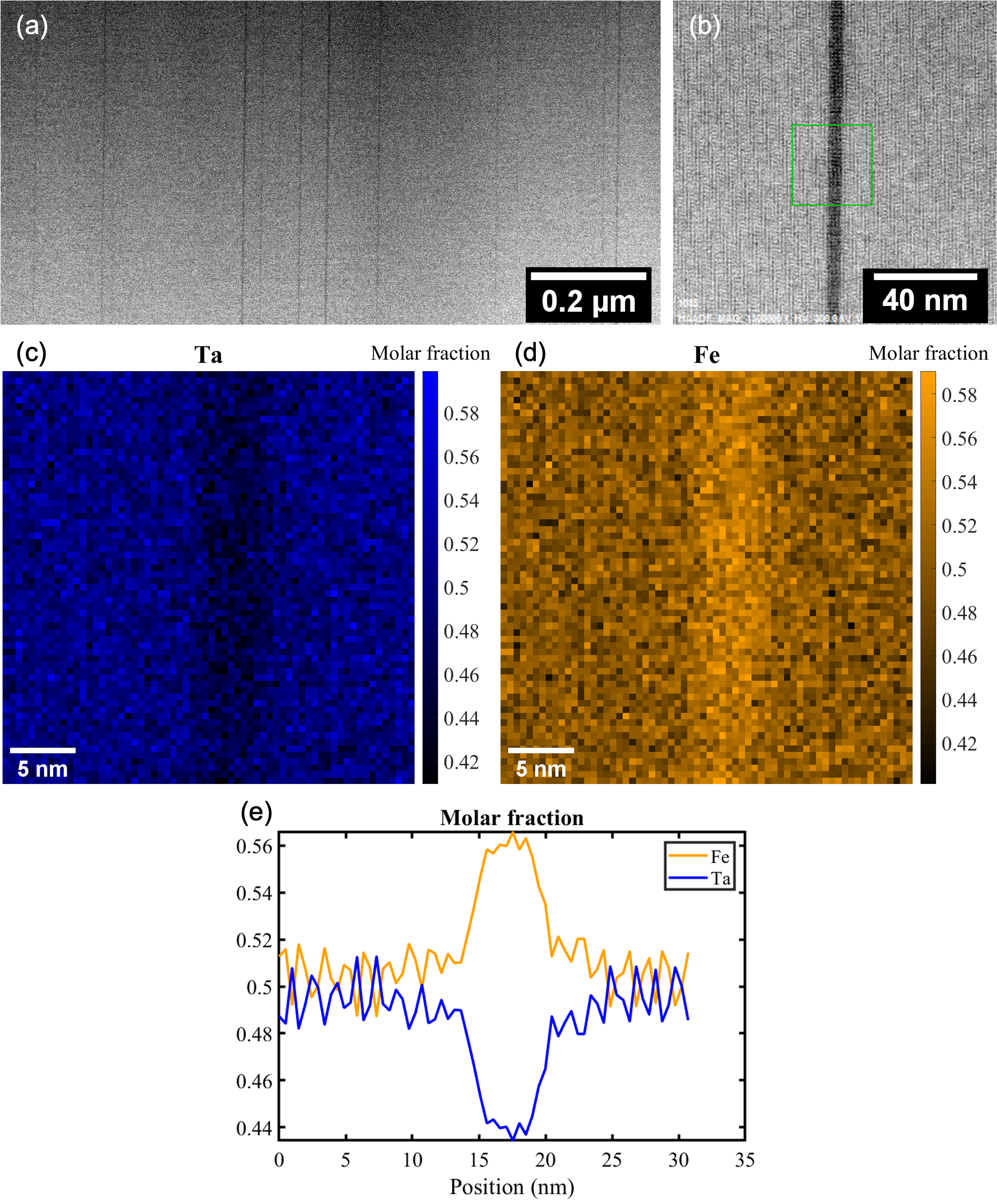}
    \caption{HR-STEM EDS measurements on the 50 at.\% Ta $\mu$-phase sample. (a) Overview of the STEM lamella showing several darker lines. In (b), one of them is displayed with higher magnification. The green square in (b) highlights the area measured by EDS. The EDS maps of the (c) Ta and (d) Fe distribution reveal an Fe-rich precipitate, as also summarised in the plot of molar fraction vs. position in (e).}
    \label{fig:3}
\end{figure*}

\section{Results}

\subsection{Experiments}

\subsubsection{EDS and EBSD analysis}

The overall chemical compositions of the combined EDS and EBSD maps of randomly selected areas of the Ta-Fe $\mu$-phase samples, were found to be 45.0, 48.8, 53.3 and 57.2 at.\% Ta, respectively, with Fe balancing them to 100 \%. The deviation from the targeted compositions are less than 1.2 at.\%, which are typical errors for EDS measurements. We have hence verified successful sample preparation for systematically investigating the influence of a gradually increasing Ta content, compare \cite{gasper2024preparation}. Furthermore, the EBSD measurements reveal the presence of several straight boundaries, ranging from a few hundred $\mu$m to several mm in length, which can be identified as twin structures. A closer examination of the IPF of the $\mu$-phase samples with different compositions, shown in Figure \ref{fig:1}, suggests that the Ta content influences both the twins' structure and the frequency of their occurrence. To further investigate this observation, the orientation relationship of the twins was studied in greater detail.

In the first two $\mu$-phase samples with target compositions of 46 and 50 at.\% Ta, only basal twins were observed. With the basal plane acting as the twin boundary plane, the orientation difference between the parent and twin grains is a 60° (or 180°) rotation along the $c$-axis, resulting in a different third Euler angle (Bunge), while the first two Euler angles remain nearly identical for both grains. As the Ta content of the Ta-Fe $\mu$-phase increases, the twin orientation relationship and, therefore, the twin boundary plane change. In the case of the $\mu$-phase samples with 54 and 58 at.\% Ta, the pyramidal $\{1\bar{1}02\}$ twin appears dominant, although some basal twins were also observed, as shown in an additional IPF map of the Ta$_7$Fe$_6$ $\mu$-phase sample provided in Figure S7.

To confirm these observations and study the atomic structures of the twin boundaries in more detail, STEM lamellae of both identified twin types were prepared and investigated by HR-STEM. The location of the lamella corresponding to the basal twin in the $\mu$-phase sample with 50 at.\% Ta is shown in Figure \ref{fig:2} (a) and the location of the pyramidal $\{1\bar{1}02\}$ twin in the $\mu$-phase sample with 54 at.\% Ta is shown in Figure \ref{fig:2} (b).

\subsubsection{HR-STEM analysis}

\begin{figure*}[hbt!]
    \centering
    \includegraphics[width=1\linewidth]{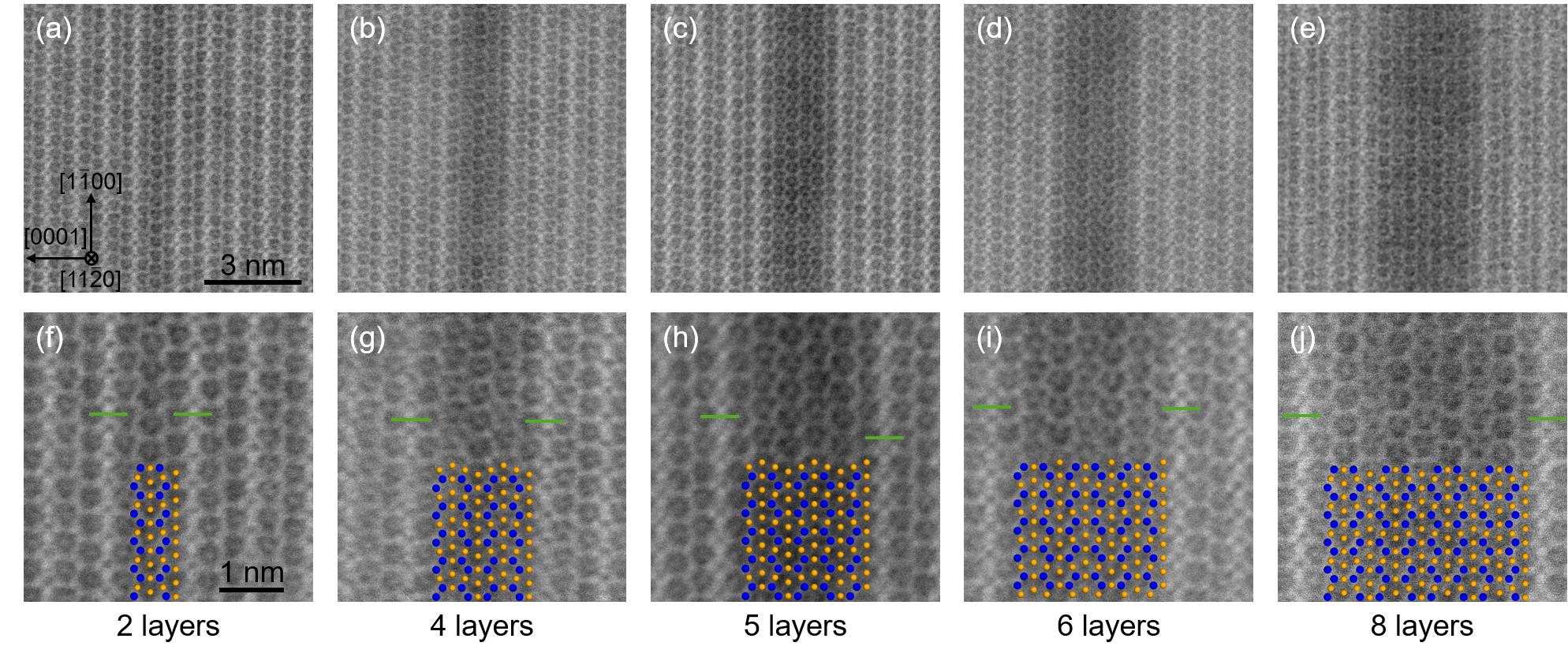}
    \caption{Varying numbers of TaFe$_2$ Laves phase layers in the 50 at.\% Ta $\mu$-phase at different magnifications. The images show two Laves phase layers in (a) and (f), four layers in (b) and (g), five layers in (c) and (h), six layers in (d) and (i) and eight layers in (e) and (j). The atomic configurations of the TaFe$_2$ Laves phase layers are superimposed on the HAADF-STEM images. The blue spheres indicate Ta and the yellow spheres Fe atomic columns. Green markers indicate the atomic positions, illustrating an atomic displacement with an odd number of Laves layers in (h).}
    \label{fig:4}
\end{figure*}

\begin{figure*}[hbt!]
    \centering
    \includegraphics[width=1\linewidth]{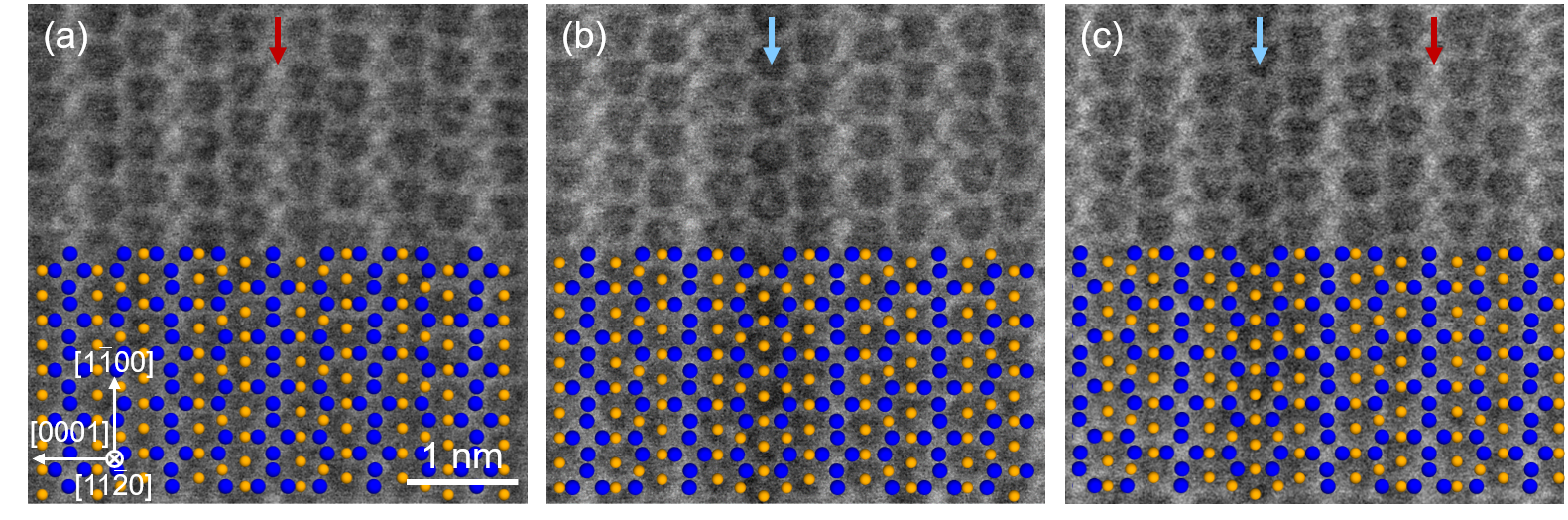}
    \caption{Basal twin boundary structures in the 50 at.\% Ta $\mu$-phase sample. (a) TB$_{\mathrm{CN15}}$ twin boundary (indicated by the red arrow), (b) TB$_\mathrm{K}$(2 Laves) twin boundary (indicated by the blue arrow) and (c) TB$_\mathrm{K}$(2 Laves) and TB$_{\mathrm{CN15}}$ twin boundaries next to each other. The atomic configurations of the basal twin boundary structures are superimposed on the HAADF-STEM images. The blue spheres indicate Ta and the yellow spheres Fe atomic columns.}
    \label{fig:5}
\end{figure*}

In the 50 at.\% Ta $\mu$-phase lamella (compare Figure \ref{fig:2} (a)) several darker lines were observed in the HAADF-STEM micrograph. These could be characterised as Fe-rich precipitates by EDS measurement, see Figure \ref{fig:3}. 
Upon examining their atomic ordering, these Fe-rich precipitates with different widths were identified as C14 TaFe$_2$ Laves phase layers. Within the lamella, regions containing two, four, five, six and eight adjacent TaFe$_2$ layers were observed, with an apparent preference for even numbers of layers, see Figure \ref{fig:4}. While odd numbers of TaFe$_2$ layers induce atomic displacement in the $\mu$-phase matrix, even numbers keep consistent atomic positions, as highlighted by the green markers in Figure \ref{fig:4}. Further analysis of the atomic ordering in the $\mu$-phase next to the Laves phase layers revealed that these layers can also act as twin boundaries, e.g., for two and five Laves phase layers in Figure \ref{fig:4} (a) and (c), respectively. In the following, this basal twin boundary structure will be referred to as TB$_\mathrm{K}$(2 Laves).

Furthermore, the middle atomic layer of the Zr$_4$Al$_3$ building block (CN15) was identified as an alternative twin boundary, which will be designated as TB$_{\mathrm{CN15}}$ (see Figure \ref{fig:5} (a)). For comparison, a TB$_\mathrm{K}$(2 Laves) twin boundary, as well as both identified basal twin boundary types, TB$_\mathrm{K}$(2 Laves) and TB$_{\mathrm{CN15}}$, are shown adjacent to each other in Figure \ref{fig:5} (b) and (c), respectively. In the STEM lamella of the 50 at.\% Ta $\mu$-phase, several more twin boundaries, including TB$_\mathrm{K}$(2 Laves) and TB$_{\mathrm{CN15}}$ were observed. As a result, HR-STEM revealed a broader variety of basal twin boundary structures that share the same macroscopic characteristics as those identified by EBSD, yet could not be distinguished by EBSD alone.

\begin{figure*}[hbt!]
    \centering
    \includegraphics[width=1\linewidth]{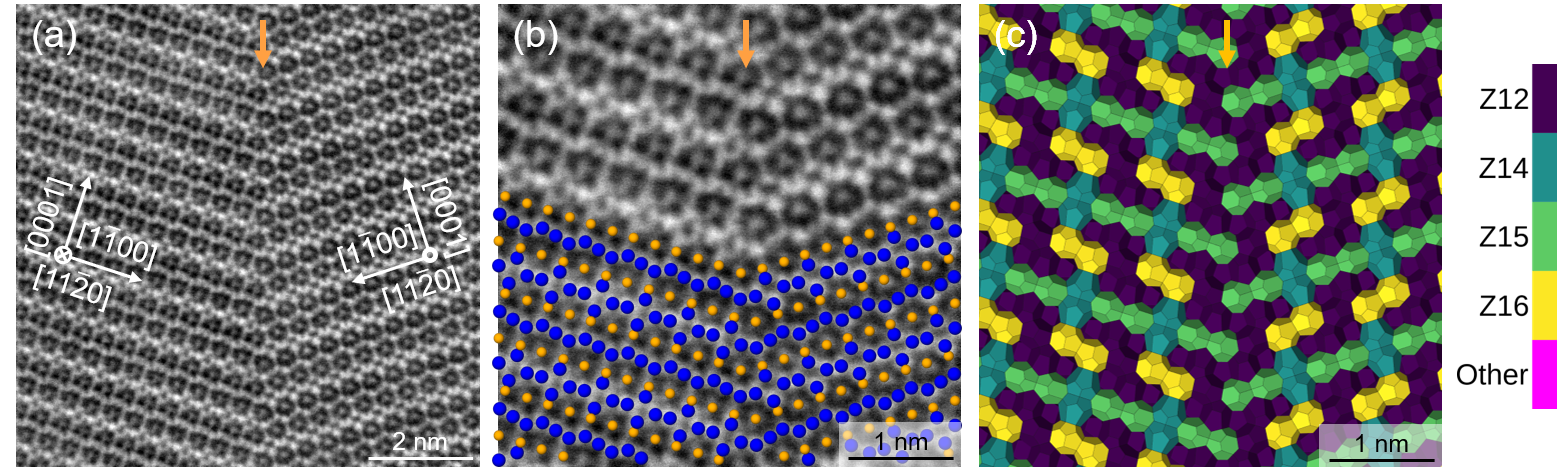}
    \caption{Pyramidal $\{1\bar{1}02\}$ twin boundary in the 54 at.\% Ta $\mu$-phase. HAADF-STEM images in (a) and (b) of different magnifications, with the atomic model of the twin boundary structure overlaid in (b). The blue spheres indicate Ta and the yellow spheres Fe atomic columns. (c) Atomic packing of the pyramidal $\{1\bar{1}02\}$ twin boundary analysed using Voronoi tessellation. Meshes coloured based on the coordination numbers of Frank-Kasper polyhedra.}
    \label{fig:6}
\end{figure*}

The investigation of the 54 at.\% Ta $\mu$-phase sample confirmed the pyramidal $\{1\bar{1}02\}$ twin boundary plane, as assumed from the EBSD orientation relationship. Only this single twin, shown in Figure \ref{fig:6}, could be observed in the prepared STEM lamella. In the images in Figure \ref{fig:6}, the twin boundary is located in the centre, perpendicular to the lamella image plane. The middle $3a$ site of the triple layer, which is occupied by a smaller B atom for the stoichiometric A$_6$B$_7$ prototype $\mu$-phase, is filled with a larger A atom in case of the A$_7$B$_6$ composition, compare Figure \ref{fig:1}. The investigated $\mu$-phase sample has a target Ta$_7$Fe$_6$ composition.

\subsection{DFT calculations}

\begin{figure*}[hbt!]
    \centering
    \includegraphics[width=1\linewidth]{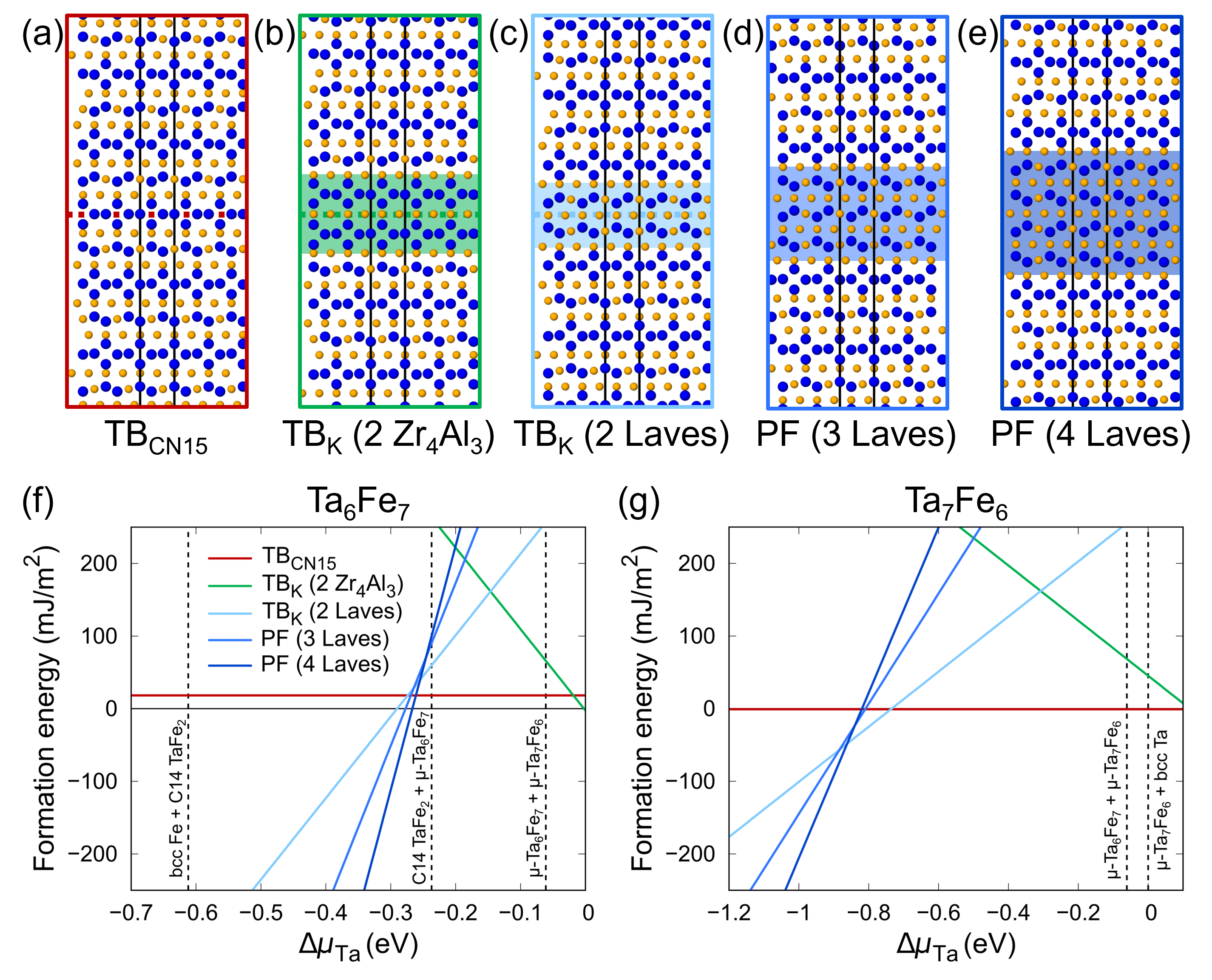}
    \caption{DFT calculations of the formation energies of twin boundaries and planar faults in the Ta$_6$Fe$_7$ and the Ta$_7$Fe$_6$ $\mu$-phases as a function of the Ta chemical potential $\Delta \mu_\text{Ta}$. Atomic configurations of (a) TB$_{\mathrm{CN15}}$, (b) TB$_\mathrm{K}$(2 Zr$_4$Al$_3$), (c) TB$_\mathrm{K}$(2 Laves), (d) PF(3 Laves), and (e) PF(4 Laves) in the Ta$_6$Fe$_7$ $\mu$-phase. The blue spheres indicate Ta and the yellow spheres Fe atomic columns. Dashed lines indicate twin boundaries, while coloured boxes highlight the faulted layers in the $\mu$-phase structure. Metastable defect phase diagrams of aforementioned twin boundaries and planar faults for the (f) Ta$_6$Fe$_7$ and the (g) Ta$_7$Fe$_6$ $\mu$-phases. Vertical dashed lines indicate the various equilibrium limits previously listed in Table \ref{tab:boundaries}. In the Ta$_6$Fe$_7$ phase, an intermediate equilibrium limit between Ta$_6$Fe$_7$ and C14 TaFe$_2$ constrained within $a$ and $b$ lattice of Ta$_6$Fe$_7$ is additionally shown.}
    \label{fig:7}
\end{figure*}

The planar defect formation was investigated across the chemical potential space via the defect phase diagram, as shown in Figure \ref{fig:7}. DFT calculations were performed to compare the chemically driven formation of various planar defects, including TB$_{\mathrm{CN15}}$, TB$_\mathrm{K}$(2 Zr$_4$Al$_3$), TB$_\mathrm{K}$(2 Laves), PF(3 Laves) and PF(4 Laves), between the Ta$_6$Fe$_7$ and the Ta$_7$Fe$_6$ $\mu$-phases. Figure \ref{fig:7} (a-e) presents the atomic structures of these planar defects, as detailed in the methods section. In the corresponding planar defects in Ta$_7$Fe$_6$, all $3a$ sites within the triple layers are fully substituted by Ta. The defect formation energy represents the energy required to generate a planar defect relative to the pristine (host) structure at a given chemical potential.

The composition of the defect structure determines the slope of its formation energy as a function of the Ta chemical potential. Defects with a higher Ta content become increasingly favourable as the Ta chemical potential increases, while defects with a higher Fe content are increasingly favourable when the Ta chemical potential decreases. Twin boundaries TB$_{\mathrm{CN15}}$ do not alter composition, therefore exhibit constant formation energies across the chemical potential range and appear as horizontal lines in the defect phase diagrams. In both Ta$_6$Fe$_7$ and Ta$_7$Fe$_6$ $\mu$-phases, twin boundaries and planar faults containing additional Laves phase layers exhibit a positive slope, indicating decreased preference with increasing Ta chemical potential. In contrast, the TB$_\mathrm{K}$(2 Zr$_4$Al$_3$) structure, which contains an additional Zr$_4$Al$_3$ building block, exhibits a negative slope, signifying greater preference at higher Ta chemical potential. Another common feature in both Ta$_6$Fe$_7$ and Ta$_7$Fe$_6$ defect phase diagrams (Figure \ref{fig:7} (f) and (g)) is the low formation energy of TB$_{\mathrm{CN15}}$, calculated as 18.3 and -0.5 mJ/m$^2$, respectively. The slightly negative formation energy of TB$_{\mathrm{CN15}}$ in Ta$_7$Fe$_6$ lies within the precision limit of DFT and becomes more positive with increasing simulation cell size (see Figure S5). 

The atomic configurations of pyramidal twin boundaries TB$_{\mathrm{Pyramidal}}$ were constructed based on the HR-STEM observations. However, the number of atoms in these systems exceeds the practical limits for DFT calculations without artificially applying constraints. Therefore, we only analysed the TB$_{\mathrm{Pyramidal}}$ structures using the Voronoi tessellation. Notably, the atomic packing of TB$_{\mathrm{Pyramidal}}$ follows the TCP packing rule where all local atomic environments are characterised by Z12, Z14, Z15 and Z16 Frank-Kasper clusters, as illustrated in Figure \ref{fig:6} (c).

At low Ta chemical potential, planar faults with larger volumes of Laves phase units possess lower formation energies than those with fewer layers in Ta$_6$Fe$_7$, following the trend: $\Delta E_F^\text{PF(4 Laves)}$ $<$ $\Delta E_F^\text{PF(3 Laves)}$ $<$ $\Delta E_F^\text{TB$_\text{K}$(2 Laves)}$. This trend aligns with the higher stability of the Laves phase at lower Ta content, as its stoichiometric structure contains 33.3 at.\% Ta. As the chemical potential of Ta increases, the formation energy ordering of Laves-containing planar defects changes at the chemical potential slightly below the formation of constraint C14 TaFe$_2$ in Ta$_6$Fe$_7$ ($\Delta \mu_\text{Ta}$ = -0.24 eV), as shown in Figure \ref{fig:7} (g). Beyond this intersection, the stability order switches to: $\Delta E_F^\text{PF(2 Laves)}$ $<$ $\Delta E_F^\text{PF(3 Laves)}$ $<$ $\Delta E_F^\text{TB$_\text{K}$(4 Laves)}$. This indicates that the formation of thicker Laves phase faults becomes less favourable as the availability of Ta increases and the one of Fe decreases. The dominant chemical potential window of TB$_\mathrm{K}$(2 Zr$_4$Al$_3$) is beyond the upper thermodynamic limit of the Ta chemical potential in Ta$_6$Fe$_7$, rendering it inaccessible. In Ta$_7$Fe$_6$, planar defects containing additional Ta-rich Laves phase units are thermodynamically unfavourable. The formation energies of TB$_\mathrm{K}$(2 Laves), TB$_\mathrm{K}$(3 Laves) and TB$_\mathrm{K}$(4 Laves) lie outside the accessible chemical potential range, indicating that such defects do not favourably form in Ta$_7$Fe$_6$. The formation of TB$_\mathrm{K}$(2 Zr$_4$Al$_3$) remains unfavourable in Ta$_7$Fe$_6$. 

The interplanar distances of Laves layers in the simulated Ta$_6$Fe$_7$ structures, including TB$_\text{K}$ (2 Laves), PF(3 Laves) and PF(4 Laves), are compared with those in the pristine $\mu$ and C14 Laves phases in Figure S8 and Figure S9. Overall, the interplanar distances within the planar Laves faults fall between the corresponding values of the $\mu$-phase and the Laves phase. 
For instance, the spacing between the triple layer and the Kagomé layer (annotated as TP-Kagomé) is 1.46 $\text{\AA{}}$ in C14 TaFe$_2$ and 1.64 $\text{\AA{}}$ in $\mu$-Ta$_6$Fe$_7$. Within the planar fault region, the TP-Kagomé spacing varies with the distance from the twin boundary or the fault center: it decreases and approaches the value of the pristine Laves phases (1.46 $\text{\AA{}}$) as the Laves layer approaches the midpoint. In contrast, the spacing within the triple layer itself increases toward the fault center. Moreover, the Ta–Fe distances inside the Ta–Fe–Ta triple layers become irregular in the presence of planar defects.

\section{Discussion}

\subsection{Chemically tailored planar defects}

The formation and prevalence of planar defects in $\mu$-phases are strongly influenced by the chemical composition. The experimental investigations using EBSD and HAADF-STEM, combined with DFT calculations of defect formation energies, provide insights into the emergence of different planar defects across varying Ta chemical potential. At a low Ta content, the $\mu$-phase coexists with the C14 TaFe$_2$ Laves phase due to phase decomposition. As the Ta content increases, the $\mu$-phase becomes dominant, with basal twinning and Laves phase layers emerging as the primary planar defects. HAADF-STEM imaging (Figures \ref{fig:4} and \ref{fig:5}) reveals C14 TaFe$_2$ Laves phase lamellae within the $\mu$-phase, such as forming TB$_\mathrm{K}$(2 Laves) twin boundaries. These findings are rationalised by the DFT calculations, where the formation of Laves phase layers becomes exothermic at low Ta chemical potential. A similar planar fault structure has been reported in Fe-rich Mo-Fe $\mu$-phases via HR-TEM, where the missing Zr$_4$Al$_3$ building blocks lead to the widespread Laves phase layers as dominant basal planar faults \cite{schroders2019structure}. Similar compositional effects on stability or metastability of basal and pyramidal planar defects in Nb-rich C14 NbFe$_2$ Laves phases, especially the presence of extended Zr$_4$Al$_3$ layers, have been investigated via HR-STEM and DFT calculations \cite{vslapakova2020atomic}. The further DFT-bonding analysis explains the preference for this type of Zr$_4$Al$_3$ planar faults and attributes it to the local reduction in charge transfer.

\begin{figure*}[hbt!]
    \centering
    \includegraphics[width=1\linewidth]{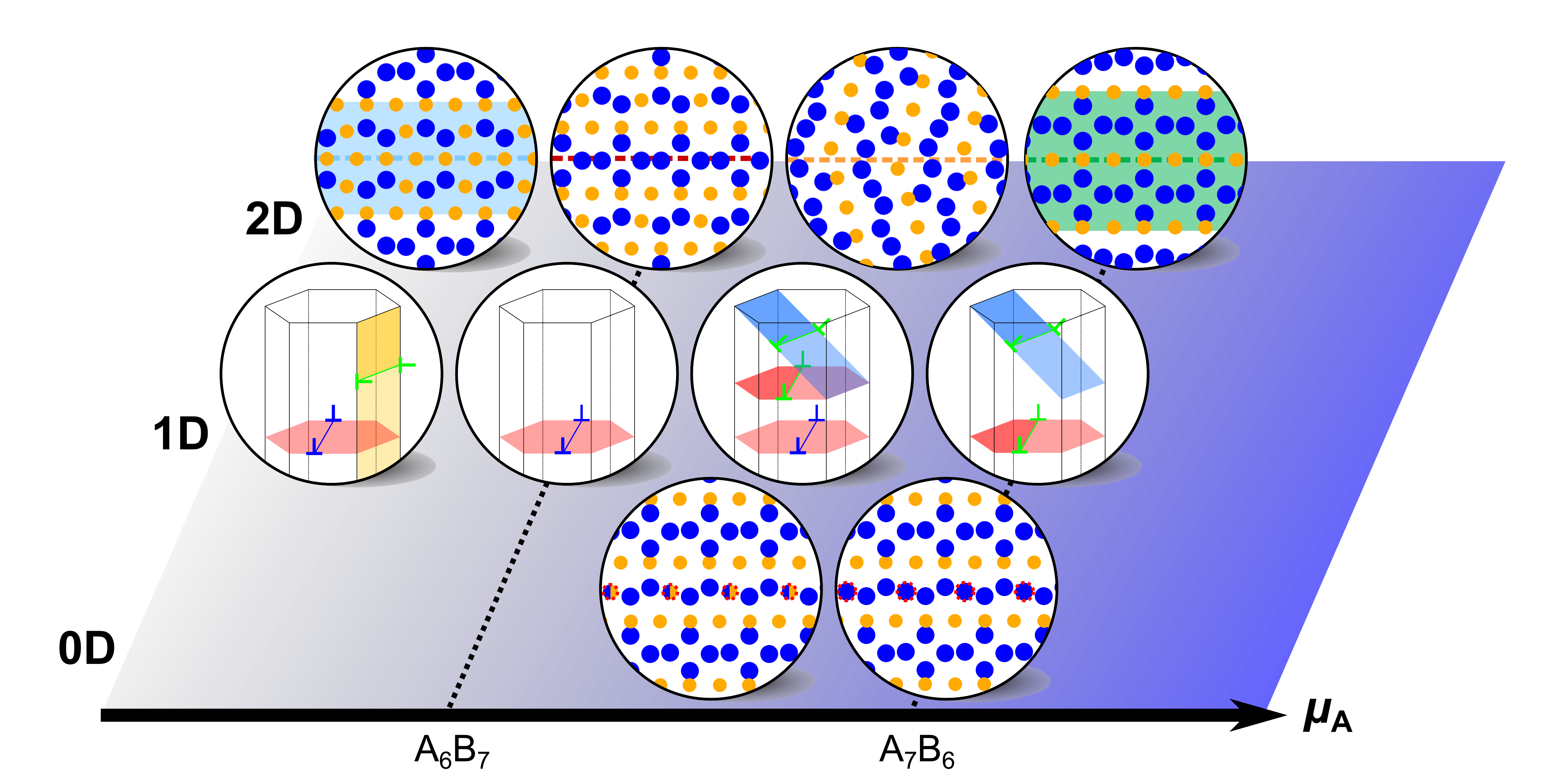}
    \caption{Schematic summary of the defect landscape in $\mu$-phases across varying chemical compositions. 0D defects: Dashed circles indicate site occupancies deviating from the stoichiometric structure. 1D defects: Blue symbols represent full dislocations, while green symbols denote partial dislocations. 2D defects: Dashed lines indicate twin boundaries, whereas coloured boxes highlight the stacking of the identical fundamental crystal building blocks in the $\mu$-phase structure. The blue spheres indicate the larger A and the yellow spheres the smaller B atomic columns.}
    \label{fig:9}
\end{figure*}

At high Ta concentrations, pyramidal twins became observable. EBSD and HAADF-STEM analyses confirmed the presence of pyramidal $\{1\bar{1}02\}$ twin boundaries, shown in Figure \ref{fig:6}. The absence of Laves phase layers suggests that their formation is thermodynamically suppressed at a higher Ta content. This is consistent with the DFT defect phase diagrams (Figure \ref{fig:7}), which indicate that Laves phase faults (TB$_\mathrm{K}$(2 Laves), PF(3 Laves) and PF(4 Laves)) become unfavourable as the Ta chemical potential increases. This transition suggests that higher Ta chemical potential promotes the formation of pyramidal twins, replacing the dominant basal defect structures at low Ta chemical potential. A comparable pyramidal twinning has been reported in a melt-spun Nb-rich Nb$_7$Ni$_6$ $\mu$-phase via HAADF-STEM \cite{zhao2023atomic}, where the chemical potential conditions closely resemble those of Ta$_7$Fe$_6$, further supporting the correlation between chemical potential and twin boundary formation. Zhao et al. attributed the twin formation to crystallographic slip via generalised stacking fault energy calculations using DFT, leading to their classification as deformation twins formed at high temperatures \cite{zhao2023atomic}. In contrast, in the present study, pyramidal twins were identified in the as-cast Ta$_7$Fe$_6$ sample, indicating that they formed during solidification, classifying them as solidification twins, rather than as a result of high-temperature deformation. Given the TCP packing structure at the pyramidal twin boundary, the associated formation energy is expected to be low. The exact pyramidal twin boundary energy requires future atomistic simulations using an accurate interatomic potential for the $\mu$-phase. To further validate the growth twinning mechanisms, future theoretical efforts are required, e.g., molecular dynamics simulations on solidification and crystal growth during annealing incorporating various crystal nuclei.

At an over-saturated Ta chemical composition (58 at.\%), the $\mu$-phase microstructure features a coexistence of pyramidal and basal twins. Experimental EBSD observations (Figure \ref{fig:1}) confirmed this, showing that while pyramidal twins remain dominant, a small fraction of basal twins is still present. This highlights the competitive nature of different planar faults, which are likely influenced by thermodynamic metastability governed by chemical potential. The basal twin boundary TB$_\mathrm{K}$(2 Zr$_4$Al$_3$), characterised by two adjacent Zr$_4$Al$_3$ building blocks, shows increasing preference with higher Ta chemical potential. Nevertheless, DFT calculations indicate that the formation of TB$_\mathrm{K}$(2 Zr$_4$Al$_3$) in the accessible chemical potential windows of Ta$_6$Fe$_7$ and Ta$_7$Fe$_6$ is endothermic. Similar basal planar faults have been reported in the Ta-rich Ta-Fe $\mu$-phase \cite{chen2025defect} and $\mu$-phase precipitates in multi-component Ni-base superalloys \cite{gao2016situ,ma2018atomic,cheng2021atomic}. However, in the multi-component systems, the site occupancies and chemical potential driving forces remain ambiguous, making direct comparisons challenging.

\subsection{Competitive defect formation in chemical potential space}

The competition among crystallographic defects formed during solidification and deformation processes plays a crucial role in determining the prevalence of specific defects. In the case of planar fault formation in $\mu$-phases during solidification, the formation of basal faults, including Laves phases layers, is thermodynamically favourable and dominates in the 50 at.\% Ta $\mu$-phase. Despite the expected low formation energy of TB$_\mathrm{Pyramidal}$, the dominance of basal planar faults may inhibit the formation and extension of pyramidal twins. In contrast, at higher Ta chemical potential, such as in the 54 at.\% Ta $\mu$-phase, the formation of Laves phase layers is completely suppressed, therefore, the formation and propagation of pyramidal twinning is no longer hindered by basal planar faults. For an even higher Ta content (58 at.\% Ta), the formation of TB$_\mathrm{K}$(2 Zr$_4$Al$_3$) becomes more thermodynamically favourable, potentially leading to the coexistence of basal and pyramidal twinning in the $\mu$-phase (see Figure \ref{fig:1}). 

Similarly, the competition among deformation-induced defects has been reported in other $\mu$-phases \cite{luo2023tailoring, luo2023plasticity, luo2024non, schroders2018room, schroders2019structure}. In off-stoichiometric Nb$_7$Co$_6$, additional Nb atoms fully occupy the $3a$ sites within the Laves layers, as demonstrated by atomic-resolution EDS \cite{luo2023tailoring}, consistent with DFT results on Nb-Ni $\mu$-phases \cite{sluiter2003site}. These chemically driven site occupancies significantly influence dislocation-mediated plastic deformation and, consequently, the mechanical properties of the material \cite{luo2023tailoring, luo2023plasticity}. Specifically, the critical resolved shear stress (CRSS) in micropillar compression and nanoindentation hardness decrease substantially when the $3a$ sites are fully occupied by Nb. As the Nb content increases from Nb$_6$Co$_7$ to Nb$_7$Co$_6$, the dominant dislocation mechanism transitions from crystallographic slip governed by full dislocations to a glide-and-shuffle mechanism dominated by partial dislocations and stacking faults, which has been identified by conventional TEM and HAADF-STEM and rationalized using DFT-based transition state tools \cite{luo2023tailoring,luo2023plasticity}. For the intermediate composition containing 49.5 at.\% Nb, both full and partial dislocations have been observed, indicating a coexistence of multiple deformation mechanisms. Analogously, in Ta–Fe $\mu$-phases, this chemically driven transition in basal dislocation mechanisms is expected. Furthermore, the presence and density of Laves-phase layers are anticipated to affect their mechanical properties, since the Laves phase is intrinsically stiffer than the $\mu$-phase matrix.

The interactions between crystallographic defects during deformation processes can also suppress the prevalence of specific defects and moderate their overall contribution to plasticity. One example is the reaction between non-basal 0.07$\langle \bar{5}502 \rangle\{1\bar{1}05\}$ partial dislocations and basal stacking faults in Nb-Co (49.5 at.\% Nb) and Ta-Fe (54 at.\% Ta) $\mu$-phases \cite{luo2024non}. Despite its high Peierls stress, the $\{1\bar{1}05\}$ partial dislocation remains glissile at room temperature, as demonstrated by its activation during micropillar compression and nanoindentation. However, the widespread presence of basal stacking faults impedes the propagation of non-basal dislocations, leading to stress concentration at their intersections and, consequently, crack initiation. As a result, the occurrence of $\{1\bar{1}05\}$ partial dislocations and stacking faults is significantly suppressed. A similar phenomenon has been observed in Fe-rich Mo-Fe $\mu$-phase, where the absence of Zr$_4$Al$_3$ building blocks results in the widespread formation of Laves phase layers as dominant planar faults \cite{schroders2018room,schroders2019structure}. In the Mo-Fe $\mu$-phase, pyramidal plasticity is largely suppressed, while basal and prismatic dislocations primarily mediate plasticity. 

Building upon the findings of this study and previous investigations on dominant defect structures across different chemical potential windows, a defect landscape in $\mu$-phases illustrates the competing defect phases \cite{korte2022defect, zhou2025} in Figure \ref{fig:9}. This landscape maps the distribution of 0D site occupancy variations, 1D dislocation types, and 2D planar faults within the chemical potential space. The defect landscape highlights a unique opportunity for tailoring the mechanical and functional properties of $\mu$-phases through defect engineering, achieved by modulating the chemical composition. 
The evolution of interplanar spacings — becoming more Laves-phase-like near the centre of a Laves-phase planar fault and more $\mu$-phase-like toward the  — suggests a degree of predictability in plastic deformation: Laves-like in defective regions, $\mu$-phase-like in the bulk. Correspondingly, different planar faults are expected to exhibit varying cleavage energies, influencing the fracture behaviour of $\mu$-phases in distinct ways.

Beyond mechanical properties, thermal conductivity can be deliberately tuned by controlling planar defect types (e.g., misorientation angles and solute segregation at grain boundaries in PbTe \cite{wu2023strong}) and defect densities (e.g., twin boundary density in Si \cite{Isotta2024} and Mg$_2$Si \cite{li2019dramatically}, stacking fault density in AgSbTe$_2$ \cite{Abdellaoui2019} ) via chemical potential control, as demonstrated in other intermetallics and compounds.

\section{Conclusions}
This study demonstrates that the formation and prevalence of planar defects in Ta-Fe $\mu$-phases are strongly influenced by chemical composition, as evidenced by experimental investigations using EBSD and HAADF-STEM, combined with DFT-calculated defect phase diagrams. The key findings are summarised as follows:

\begin{itemize}
    \item In the two-phase region of the C14 Laves and $\mu$-phases at 46 at.\% Ta, basal twins appear to be the dominant planar faults in the $\mu$-phase.
    \item At 50 at.\% Ta, basal twin boundaries and planar faults containing C14 TaFe$_2$ Laves phase layers are dominant in the $\mu$-phase, with their formation being thermodynamically favourable at low Ta chemical potential.
    \item In Ta$_7$Fe$_6$ (54 at.\% Ta), a transition from basal to pyramidal $\{1\bar{1}02\}$ twinning occurs, with the suppression of Laves phase layers at higher Ta chemical potential. The atomic structure of TB$_\mathrm{Pyramidal}$ follows the topologically close-packing, suggesting a low formation energy.
    \item In the 58 at.\% Ta sample, prevalent pyramidal twins coexist with some basal twins in the $\mu$-phase.
    \item The competition among crystallographic defects during solidification and deformation leads to distinct defect landscapes across varying chemical potentials, underscoring the potential for tailoring material properties via chemically driven defect engineering.
\end{itemize} 

\section*{Declaration of competing interest}
The authors declare that they have no known competing financial
interests or personal relationships that could have appeared to influence
the work reported in this paper.

\section*{Acknowledgments}
This project has received funding from the European Research Council (ERC) under the European Union’s Horizon 2020 Research and Innovation Programme (Grant Agreement No. 852096 FunBlocks). 
Funding by the Deutsche Forschungsgemeinschaft (DFG) in the SFB1394 "Structural and chemical atomic complexity – from defect phase diagrams to material properties" (Project ID 409476157) is gratefully acknowledged. Z.X. acknowledges financial support funded by the DFG – Projektnummer 562592407. The authors gratefully acknowledge the computing time provided to them at the NHR Center NHR4CES at RWTH Aachen University (project numbers p0020330). The data used in this publication was managed using the research data management platform Coscine with storage space granted by the Research Data Storage (RDS) of the DFG and Ministry of Culture and Science (MKW) of the State of North Rhine-Westphalia (DFG: INST222/1261-1 and MKW: 214-4.06.05.08 - 139057). For the management of the research data, the authors gratefully acknowledge the support of Khalil Rejiba.

\section*{Data availability}
Data will be made available on request.

\bibliographystyle{elsarticle-num}
\bibliography{cite}

\begin{thebibliography}{10}
\expandafter\ifx\csname url\endcsname\relax
  \def\url#1{\texttt{#1}}\fi
\expandafter\ifx\csname urlprefix\endcsname\relax\def\urlprefix{URL }\fi
\expandafter\ifx\csname href\endcsname\relax
  \def\href#1#2{#2} \def\path#1{#1}\fi

\bibitem{sauthoff1995intermetallics}
G.~Sauthoff, {Intermetallics}, VCH Verlagsgesellschaft mbH, 1995.

\bibitem{ferro2011intermetallic}
R.~Ferro, A.~Saccone, {Intermetallic chemistry}, Vol.~13, Elsevier, 2011.

\bibitem{pearson1972crystal}
W.~B. Pearson, {The crystal chemistry and physics of metals and alloys}, Wiley-Inter-science 135 (1972).

\bibitem{tammann1923metallographische}
G.~Tammann, K.~Dahl, {Metallographische Mitteilungen aus dem Institut f{\"u}r physikalische Chemie der Universit{\"a}t G{\"o}ttingen. CXI. {\"U}ber die Spr{\"o}digkeit metallischer Verbindungen}, Zeitschrift f{\"u}r anorganische und allgemeine Chemie 126~(1) (1923) 104--112.

\bibitem{fleischer1989intermetallic}
R.~Fleischer, D.~Dimiduk, H.~Lipsitt, {Intermetallic compounds for strong high-temperature materials; Status and potential}, Annual Review of Materials Science (United States) 19 (1989).

\bibitem{westbrook2000basic}
J.~H. Westbrook, R.~L. Fleischer, {Basic mechanical properties and lattice defects of intermetallic compounds}, John Wiley \& Sons. Inc., 2000.

\bibitem{Belin2010mechanical}
E.~Belin-Ferré, {Mechanical Properties of Complex Intermetallics}, World Scientific Publishing Company, 2010.

\bibitem{magneli1938rontgenuntersuchung}
A.~Magneli, A.~Westgren, {R{\"o}ntgenuntersuchung von {K}obalt--{W}olframlegierungen}, Zeitschrift f{\"u}r anorganische und allgemeine Chemie 238~(2-3) (1938) 268--272.

\bibitem{forsyth1962structure}
J.~Forsyth, L.~D'Alte~da Veiga, {The structure of the $\mu$-phase {C}o$_{7}${M}o$_{6}$}, Acta Crystallographica 15~(6) (1962) 543--546.

\bibitem{kumar1998sublattice}
K.~H. Kumar, I.~Ansara, P.~Wollants, {Sublattice modelling of the $\mu$-phase}, Calphad 22~(3) (1998) 323--334.

\bibitem{frank1959complex}
F.~C. Frank, J.~S. Kasper, {Complex alloy structures regarded as sphere packings. {II}. Analysis and classification of representative structures}, Acta Crystallographica 12~(7) (1959) 483--499.

\bibitem{wilson1960crystal}
C.~Wilson, D.~Thomas, F.~Spooner, {The crystal structure of {Z}r$_{4}${A}l$_{3}$}, Acta Crystallographica 13~(1) (1960) 56--57.

\bibitem{andersson1978structures}
S.~Andersson, {Structures related to the $\beta$-tungsten or {C}r$_{3}${S}i structure type}, Journal of Solid State Chemistry 23~(1-2) (1978) 191--204.

\bibitem{sinha1972topologically}
A.~K. Sinha, {Topologically close-packed structures of transition metal alloys}, Progress in Materials Science 15~(2) (1972) 81--185.

\bibitem{joubert2004mixed}
J.-M. Joubert, N.~Dupin, {Mixed site occupancies in the $\mu$ phase}, Intermetallics 12~(12) (2004) 1373--1380.

\bibitem{luo2023tailoring}
W.~Luo, Z.~Xie, S.~Zhang, J.~Gu{\'e}nol{\'e}, P.-L. Sun, A.~Meingast, A.~Alhassan, X.~Zhou, F.~Stein, L.~Pizzagalli, et~al., {Tailoring the Plasticity of Topologically Close-packed Phases via the Crystals’ Fundamental Building Blocks}, Advanced materials 35~(24) (2023) 2300586.

\bibitem{cieslak2014structural}
J.~Cieslak, J.~Przewoznik, S.~Dubiel, {Structural and electronic properties of the $\mu$-phase {F}e--{M}o compounds}, Journal of alloys and compounds 612 (2014) 465--470.

\bibitem{shoemaker1971tetraedrisch}
C.~B. Shoemaker, D.~P. Shoemaker, {Tetraedrisch dicht gepackte {S}trukturen von {L}egierungen der {\"U}bergangsmetalle}, Monatshefte f{\"u}r Chemie/Chemical Monthly 102~(6) (1971) 1643--1666.

\bibitem{luo2023plasticity}
W.~Luo, Z.~Xie, P.-L. Sun, J.-L. Gibson, S.~Korte-Kerzel, {Plasticity of the {N}b-rich $\mu$-Co$_{7}$Nb$_{6}$ phase at room temperature and 600° C}, Acta Materialia 246 (2023) 118720.

\bibitem{stollenwerk2025beyond}
T.~Stollenwerk, P.~C. Huckfeldt, N.~Z.~Z. Ulumuddin, M.~Schneider, Z.~Xie, S.~Korte-Kerzel, {Beyond Fundamental Building Blocks: Plasticity in Structurally Complex Crystals}, Advanced Materials 37~(6) (2025) 2414376.

\bibitem{gottstein2004physical}
G.~Gottstein, S.~Korte-Kerzel, {Materialwissenschaft und {W}erkstofftechnik: {P}hysikalische {G}rundlagen}, Springer-Verlag, 2025.

\bibitem{sluiter2003site}
M.~H. Sluiter, A.~Pasturel, Y.~Kawazoe, {Site occupation in the {N}i-{N}b $\mu$ phase}, Physical Review B 67~(17) (2003) 174203.

\bibitem{luo2024non}
W.~Luo, C.~Gasper, S.~Zhang, P.~Sun, N.~Ulumuddin, A.~Petrova, Y.~Lysogorskiy, R.~Drautz, Z.~Xie, S.~Korte-Kerzel, {Non-basal plasticity in the $\mu$-phase at room temperature}, Acta Materialia 277 (2024) 120202.

\bibitem{schroders2018room}
S.~Schr{\"o}ders, S.~Sandl{\"o}bes, C.~Birke, M.~Loeck, L.~Peters, C.~Tromas, S.~Korte-Kerzel, {Room temperature deformation in the {F}e$_{7}${M}o$_{6}$ $\mu$-Phase}, International Journal of Plasticity 108 (2018) 125--143.

\bibitem{schroders2019structure}
S.~Schr{\"o}ders, S.~Sandl{\"o}bes, B.~Berkels, S.~Korte-Kerzel, {On the structure of defects in the {F}e$_{7}${M}o$_{6}$ $\mu$-Phase}, Acta Materialia 167 (2019) 257--266.

\bibitem{chen2025defect}
Z.~Chen, T.~Kishi, S.~Han, K.~Kishida, H.~Inui, Defect structures and room-temperature deformation of single crystals of the mu-phase compound {F}e$_7${T}a$_6$ investigated by micropillar compression, Acta Materialia (2025) 121552.

\bibitem{gao2016situ}
S.~Gao, Z.-Q. Liu, C.-F. Li, Y.~Zhou, T.~Jin, {In situ {TEM} investigation on the precipitation behavior of $\mu$ phase in {N}i-base single crystal superalloys}, Acta Materialia 110 (2016) 268--275.

\bibitem{ma2018atomic}
S.~Ma, X.~Li, J.~Zhang, J.~Liu, P.~Li, Y.~Zhang, H.~Jin, W.~Zhang, Y.~Zhou, X.~Sun, et~al., {Atomic arrangement and formation of planar defects in the $\mu$ phase of {N}i-base single crystal superalloys}, Journal of Alloys and Compounds 766 (2018) 775--783.

\bibitem{cheng2021atomic}
Y.~Cheng, G.~Wang, J.~Liu, L.~He, {Atomic configurations of planar defects in $\mu$ phase in {N}i-based superalloys}, Scripta Materialia 193 (2021) 27--32.

\bibitem{jin2022experimental}
H.~Jin, J.~Zhang, W.~Zhang, Y.~Zhang, S.~Ma, Y.~Du, J.~Qin, Q.~Wang, {Experimental studies and {DFT} calculations to predict atomic arrangements at twin boundaries and distribution behaviors of different solutes in complex intermetallics}, Journal of Physics and Chemistry of Solids 161 (2022) 110453.

\bibitem{chen2025mechanisms}
W.~Chen, Y.~Zhao, X.~Fei, X.~Wang, P.~Nan, Y.~Zhang, H.~Jin, B.~Ge, The mechanisms of preferential occupation and planar defect transformation in the $\mu$ phase of cobalt-based superalloys, Acta Materialia (2025) 121399.

\bibitem{zhao2023atomic}
J.~Zhao, H.~Wang, B.~Wei, {Atomic-scale structural characterization and twin formation mechanisms of $\mu$ phase within refractory {N}b{N}i alloy}, Materials Characterization 201 (2023) 112921.

\bibitem{okamoto2013fe}
H.~Okamoto, {{F}e-{T}a ({I}ron-{T}antalum)}, Journal of phase equilibria and diffusion 34 (2013) 165--166.

\bibitem{witusiewicz2011experimental}
V.~Witusiewicz, A.~Bondar, U.~Hecht, V.~Voblikov, O.~Fomichov, V.~Petyukh, S.~Rex, {Experimental study and thermodynamic re-assessment of the binary {F}e--{T}a system}, Intermetallics 19~(7) (2011) 1059--1075.

\bibitem{gasper2025mechanical}
C.~Gasper, E.~M. Soysal, N.~Ulumuddin, T.~Stollenwerk, T.~Reclik, P.-L. Sun, S.~Korte-Kerzel, {Mechanical properties and deformation mechanisms of the {C}14 Laves and $\mu$-phase in the ternary {T}a-{F}e (-{A}l) system}, Materials \& Design (2025) 113625.

\bibitem{gasper2024preparation}
C.~Gasper, I.~Gao, F.~Busch, A.~Ziemons, D.~Beckers, H.~Springer, S.~Korte-Kerzel, {Preparation of Binary and Ternary Laves and $\mu$-Phases in the {T}a-{F}e(-{A}l) System for Property Analysis at the Microscale}, Metallurgical and Materials Transactions A (2024) 1--20.

\bibitem{greenwood2010chemistry}
N.~Greenwood, A.~Earnshaw, {Chemistry of the Elements}, 2nd Edition, Elsevier, 2010.

\bibitem{zhang2018evaluation}
S.~Zhang, C.~Scheu, {Evaluation of {EELS} spectrum imaging data by spectral components and factors from multivariate analysis}, Microscopy 67~(suppl\_1) (2018) i133--i141.

\bibitem{BeLi19}
B.~Berkels, C.~H. Liebscher, {Joint non-rigid image registration and reconstruction for quantitative atomic resolution scanning transmission electron microscopy}, Ultramicroscopy 198 (2019) 49--57.

\bibitem{kresse1996efficient}
G.~Kresse, J.~Furthm{\"u}ller, {{Efficient} iterative schemes for {ab} initio total-energy calculations using a plane-wave basis set}, {Physical Review B} 54~(16) (1996) 11169.

\bibitem{kresse1993ab}
G.~Kresse, J.~Hafner, {{Ab} initio molecular dynamics for liquid metals}, {Physical Review B} 47~(1) (1993) 558.

\bibitem{kresse1999ultrasoft}
G.~Kresse, D.~Joubert, {{From} ultrasoft pseudopotentials to the projector augmented-wave method}, {Physical Review b} 59~(3) (1999) 1758.

\bibitem{perdew1996generalized}
J.~P. Perdew, K.~Burke, M.~Ernzerhof, {Generalized} gradient approximation made simple, {Physical Review Letters} 77~(18) (1996) 3865.

\bibitem{methfessel1989high}
M.~Methfessel, A.~Paxton, {High-precision} sampling for {Brillouin-zone} integration in metals, {Physical Review B} 40~(6) (1989) 3616.

\bibitem{monkhorst1976special}
H.~J. Monkhorst, J.~D. Pack, {Special} points for {Brillouin-zone} integrations, {Physical Review B} 13~(12) (1976) 5188.

\bibitem{hirel2015atomsk}
P.~Hirel, {Atomsk: A tool for manipulating and converting atomic data files}, Computer Physics Communications 197 (2015) 212--219.

\bibitem{raman1966mu}
A.~Raman, {The $\mu$ Phases}, International Journal of Materials Research 57~(4) (1966) 301--305.

\bibitem{stukowski2009visualization}
A.~Stukowski, {Visualization} and analysis of atomistic simulation data with {OVITO} - the {Open} {Visualization} {Tool}, {Modelling and Simulation in Materials Science and Engineering} 18~(1) (2009) 015012.

\bibitem{korte2022defect}
S.~Korte-Kerzel, T.~Hickel, L.~Huber, D.~Raabe, S.~Sandl{\"o}bes-Haut, M.~Todorova, J.~Neugebauer, {Defect phases--thermodynamics and impact on material properties}, International Materials Reviews 67~(1) (2022) 89--117.

\bibitem{tehranchi2024metastable}
A.~Tehranchi, S.~Zhang, A.~Zendegani, C.~Scheu, T.~Hickel, J.~Neugebauer, {Metastable defect phase diagrams as roadmap to tailor chemically driven defect formation}, Acta Materialia 277 (2024) 120145.

\bibitem{vslapakova2020atomic}
M.~{\v{S}}lap{\'a}kov{\'a}, A.~Zendegani, C.~Liebscher, T.~Hickel, J.~Neugebauer, T.~Hammerschmidt, A.~Ormeci, J.~Grin, G.~Dehm, K.~S. Kumar, et~al., Atomic scale configuration of planar defects in the {N}b-rich {C}14 {L}aves phase {N}b{F}e2, Acta Materialia 183 (2020) 362--376.

\bibitem{zhou2025}
X.~Zhou, P.~Mathews, B.~Berkels, W.~Delis, S.~Saood, A.~{Shamseldeen Ali Alhassan}, P.~Keuter, J.~M. Schneider, S.~Korte-Kerzel, S.~Sandl{\"{o}}bes-Haut, D.~Raabe, J.~Neugebauer, G.~Dehm, T.~Hickel, C.~Scheu, S.~Zhang, {Materials Design by Constructing Phase Diagrams for Defects}, Advanced Materials 37 (2025) 2402191.

\bibitem{wu2023strong}
R.~Wu, Y.~Yu, S.~Jia, C.~Zhou, O.~Cojocaru-Mir{\'e}din, M.~Wuttig, {Strong charge carrier scattering at grain boundaries of PbTe caused by the collapse of metavalent bonding}, Nature Communications 14~(1) (2023) 719.

\bibitem{Isotta2024}
E.~Isotta, S.~Jiang, R.~Bueno-Villoro, R.~Nagahiro, K.~Maeda, D.~A. Mattlat, A.~R. Odufisan, A.~Zevalkink, J.~Shiomi, S.~Zhang, C.~Scheu, G.~J. Snyder, O.~Balogun, {Heat Transport at Silicon Grain Boundaries}, Advanced Functional Materials 2405413 (2024).

\bibitem{li2019dramatically}
G.~Li, J.~He, Q.~An, S.~I. Morozov, S.~Hao, P.~Zhai, Q.~Zhang, W.~A. Goddard~III, G.~J. Snyder, {Dramatically reduced lattice thermal conductivity of {M}g$_{2}${S}i thermoelectric material from nanotwinning}, Acta Materialia 169 (2019) 9--14.

\bibitem{Abdellaoui2019}
L.~Abdellaoui, S.~Zhang, S.~Zaefferer, R.~Bueno-Villoro, A.~Baranovskiy, O.~Cojocaru-Mir{\'{e}}din, Y.~Yu, Y.~Amouyal, D.~Raabe, G.~Snyder, C.~Scheu, {Density, distribution and nature of planar faults in silver antimony telluride for thermoelectric applications}, Acta Materialia 178 (2019).

\end{thebibliography}

\clearpage
\onecolumn
\appendix

\twocolumn[ 

\vspace*{8cm} 
\begin{center}
    {\Large \textbf{Supplementary Material}} \\
    \vspace{0.5cm} 
    {\Large Chemically tailored planar defect phases in the Ta-Fe $\mu$-phase} \\
    \vspace{0.5cm} 
    {\Large Gasper, et al.}
\end{center}

\vspace{1cm} 
]

\setcounter{figure}{0}
\setcounter{table}{0}
\setcounter{equation}{0}

\renewcommand{\thefigure}{S\arabic{figure}}
\renewcommand{\thetable}{S\arabic{table}}
\renewcommand{\theequation}{S\arabic{equation}}


\begin{figure*}[hbt!]
\centering
\includegraphics[width=0.5\textwidth]{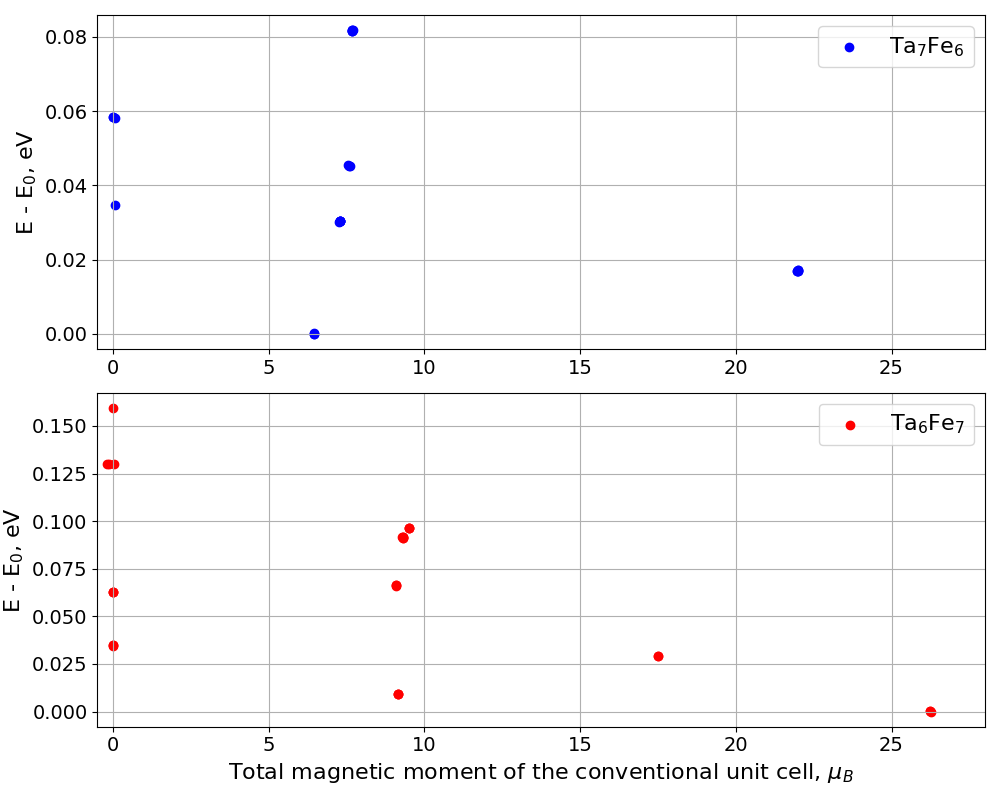}
\caption{Energy differences with respect to the lowest-energy structure among the magnetic configuration pool sampled in Ta$_7$Fe$_6$ and Ta$_6$Fe$_7$. The highest magnetic moment corresponds to the ferromagnetic configuration. }
\label{fig:energy differences}
\end{figure*}

\begin{figure*}[hbt!]
\centering
\includegraphics[width=\textwidth]{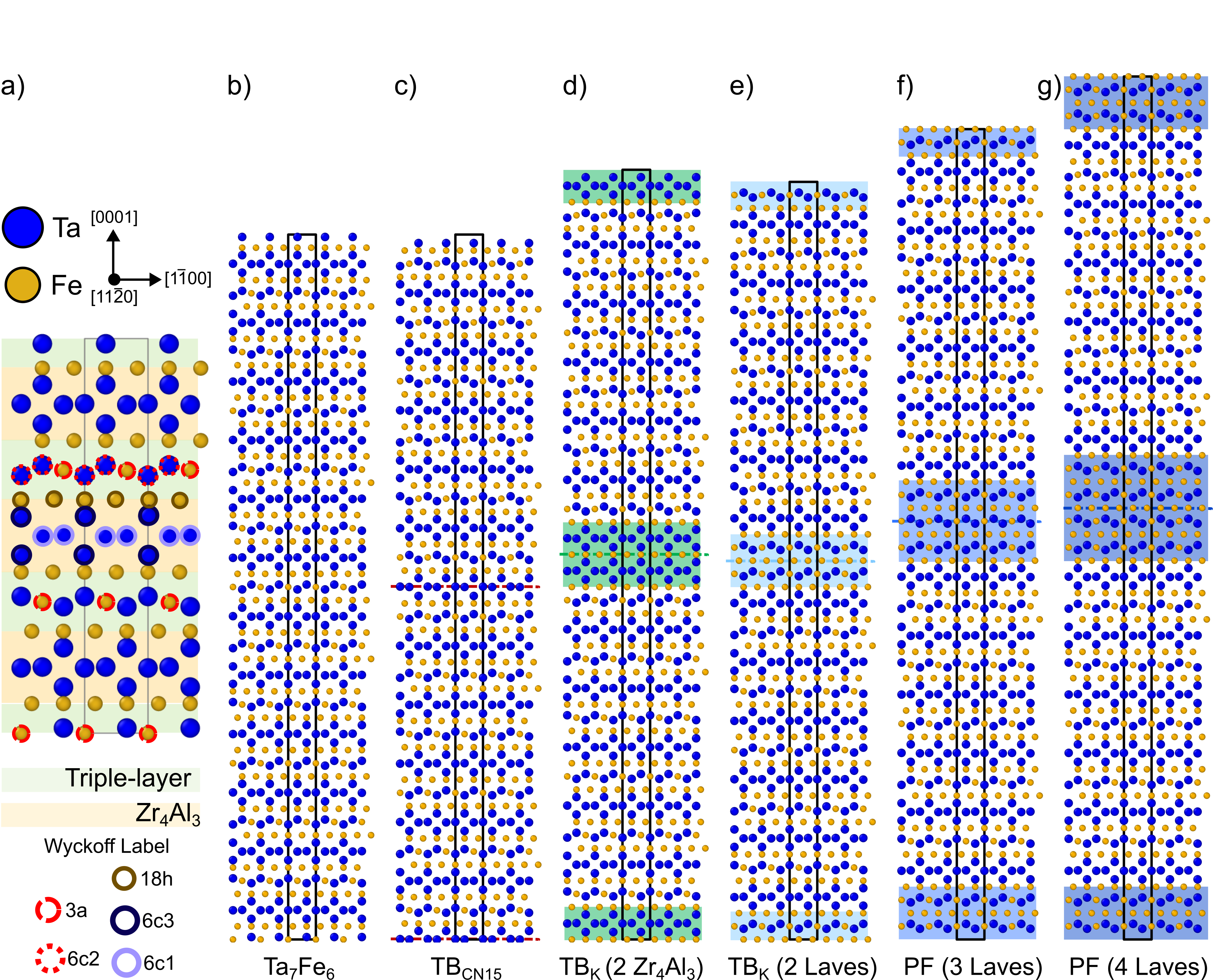}
\caption{The simulation cells of basal-oriented Ta$_6$Fe$_7$ with Ta (blue) and Fe (yellow) atoms, where (a) displays the conventional unit cell. The stacking sequence of the $\mu$-phase along the $c$-axis consists of triple layers and Zr$_4$Al$_3$ units and are highlighted in light green and light yellow, respectively. Unique Wyckoff positions of representative atoms are distinguished by different outline colors. In Ta$_6$Fe$_7$, Fe occupies the $3a$ sites within the triple layers, whereas in Ta$_7$Fe$_6$, these sites are fully occupied by Ta. The supercells used to model basal-oriented planar fault structures are presented in (b–g) for pristine Ta$_6$Fe$7$, TB$_\mathrm{{CN15}}$, TB$_K$(2 Zr$_4$Al$_3$), TB$_K$(2 Laves), PF (3 Laves), and PF (4 Laves), respectively. The planar fault segments are highlighted while the offset/twinning planes are indicated by dashed lines, as in the main figure. All schematics shown here are in composition Ta$_6$Fe$_7$. In the corresponding Ta$_7$Fe$_6$ structures, all $3a$ sites are occupied by Ta.}
\label{fig:simulation cells}
\end{figure*}


\begin{figure*}[hbt!]
\centering
\includegraphics[width=\textwidth]{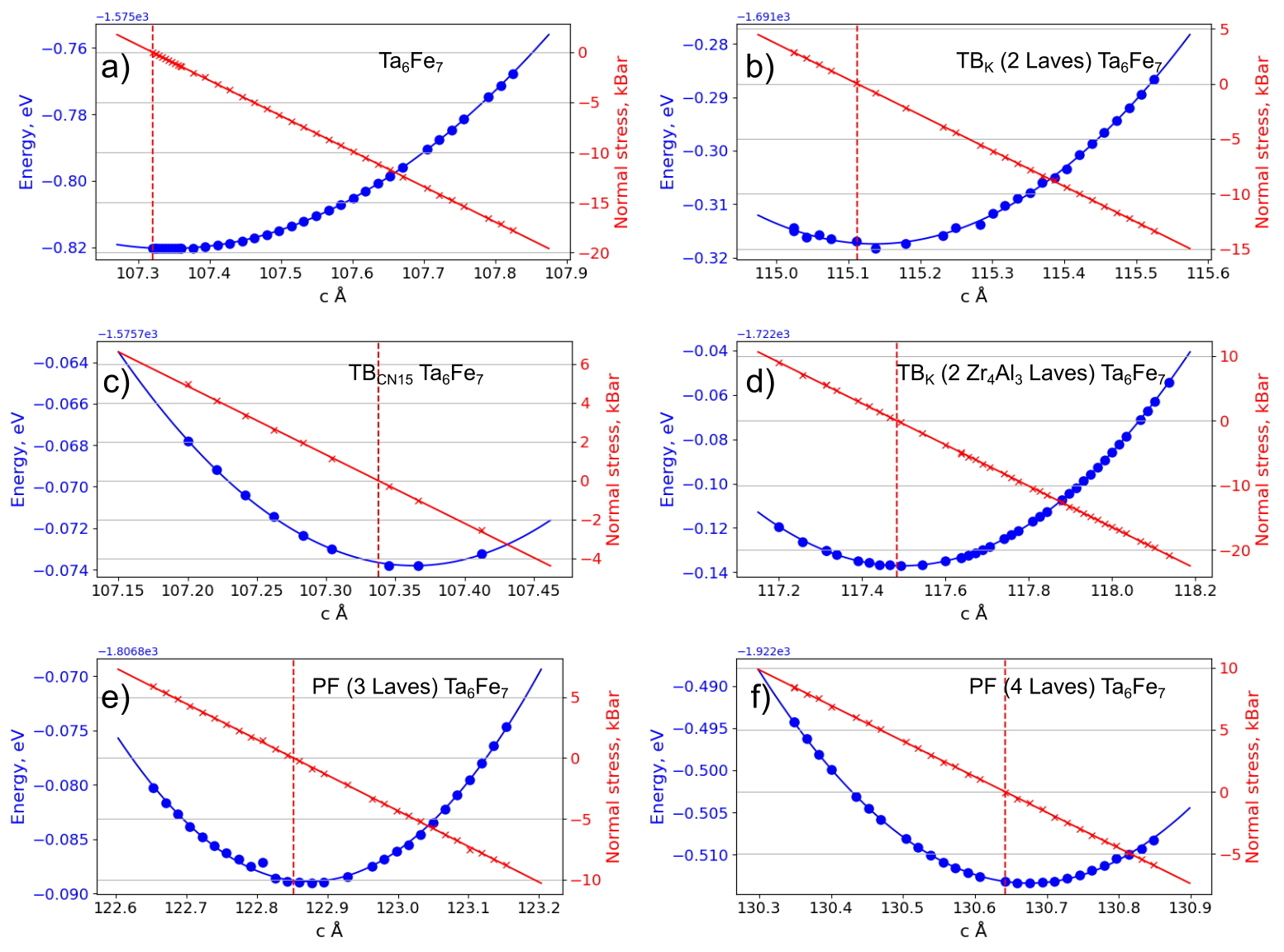}
\caption{The minimisation of the normal stress in the Ta$_6$Fe$_7$ simulation cells by scanning the dimension along the $c$-axis.}
\label{fig:minimisation 1}
\end{figure*}

\begin{figure*}[hbt!]
\centering
\includegraphics[width=\textwidth]{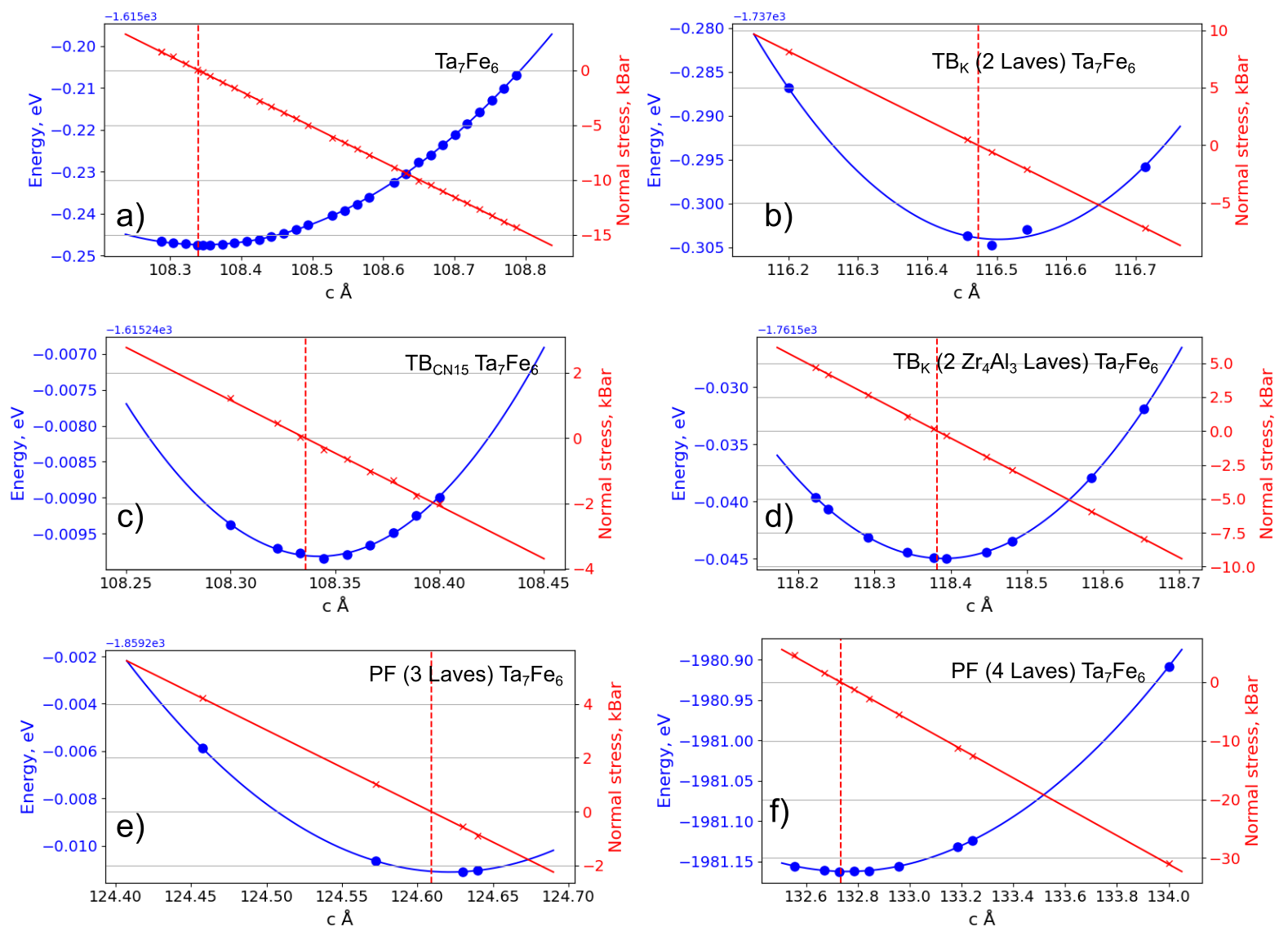}
\caption{The minimisation of the normal stress in the Ta$_7$Fe$_6$ simulation cells by scanning the dimension along the $c$-axis.}
\label{fig:minimisation 2}
\end{figure*}

\begin{figure*}[hbt!]
\centering
\includegraphics[width=\textwidth]{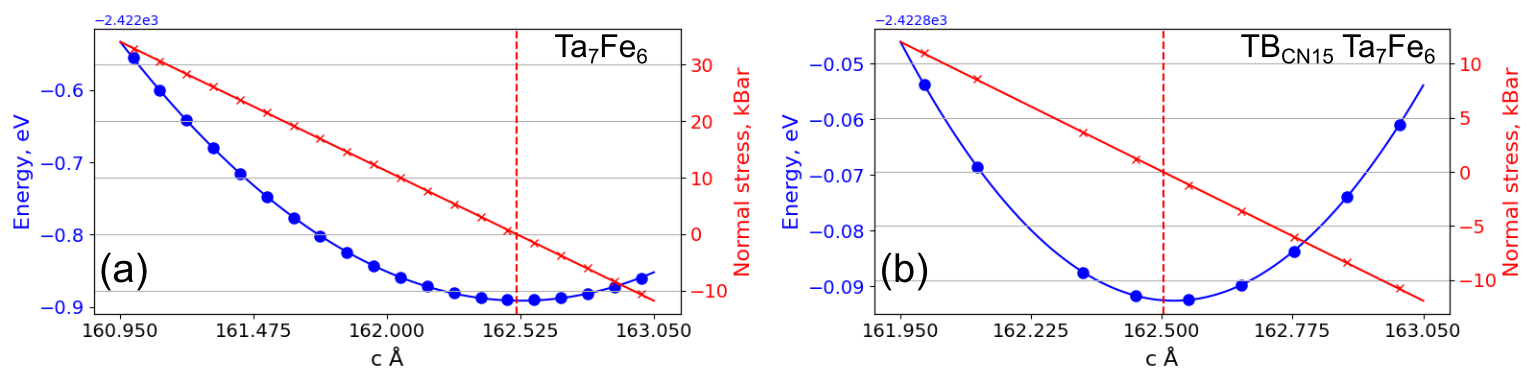}
\caption{The minimisation of the normal stress in TB$_\mathrm{CN15}$ Ta$_7$Fe$_6$ 1 $\times$ 1 $\times$ 6 supercell. The resulting defect formation energy of this structure is -0.38 mJ/m$^2$, which is 0.16 mJ/m$^2$ higher than the corresponding defect formation energy of the 1 $\times$ 1 $\times$ 4 supercell. The energy difference is within the DFT precision limit, indicating the size effects are negligible.}
\label{fig:minimisation 3}
\end{figure*}

\begin{figure*}[hbt!]
\centering
\includegraphics[width=0.5\textwidth]{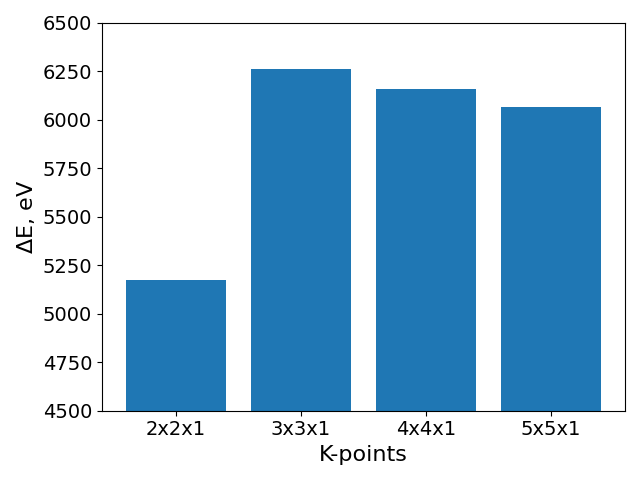}
\caption{K-point benchmarking of the basal-oriented 1 $\times$ 1 $\times$ 4 supercell using energy difference, given  $a$, $b$ and $c$  cell dimensions of 4.855, 4.855 and 54.740 $\AA$, respectively, containing 182 atoms. The k-point mesh of $5 \times 5 \times 1$ using the Monkhorst-Pack k-point scheme was used for the rest of the work, where this corresponds to a k-point per reciprocal lattice (KPPRA) of 1000.}
\label{fig:k-point}
\end{figure*}

\begin{figure*}[hbt!]
\centering
\includegraphics[width=\textwidth]{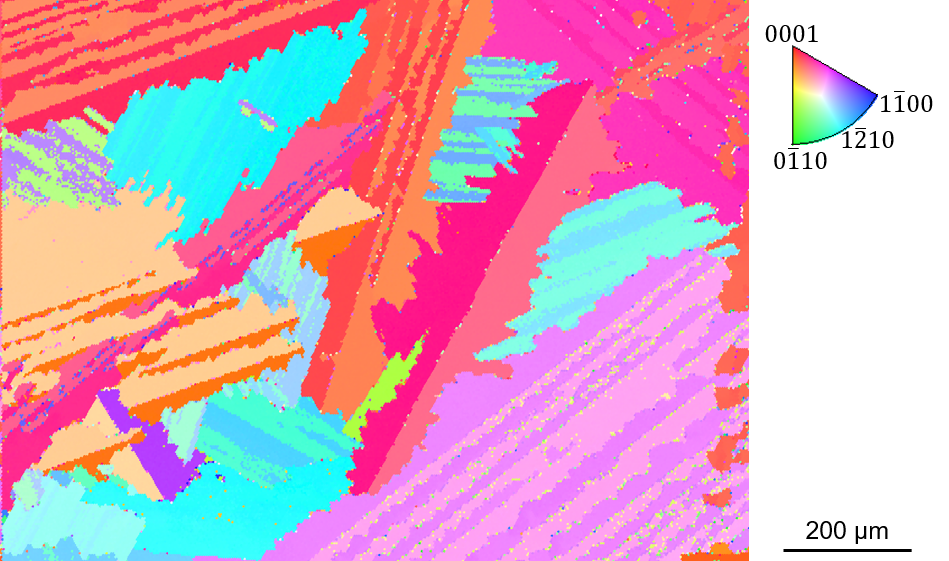}
\caption{EBSD IPF map of Ta$_7$Fe$_6$ $\mu$-phase sample with a target composition of 54 at.\% Ta and 46 at\% Fe. Apart from a larger step size of 3.5 $\mu$m, the same parameters were used as described in the methodology section of the manuscript.}
\label{fig:EBSD}
\end{figure*}

\begin{figure*}[hbt!]
\centering
\includegraphics[width=0.8\textwidth]{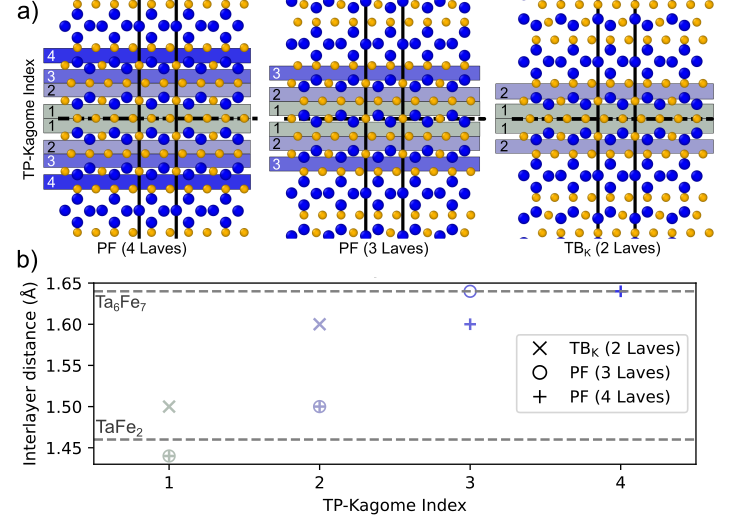}
\caption{DFT-simulated TP-Kagome interplanar distances of the Laves phase layers in PF(4 Laves), PF(3 Laves) and TB$_K$(2 Laves) of the  Ta$_6$Fe$_7$ $\mu$-phase, where each distinct TP-Kagome interlayers are indexed from 1 to 4 depending on its distance from the mirror plane. For comparison, the TP-Kagome distance of TaFe$_2$ and Ta$_6$Fe$_7$ are shown as labeled dashed lines at 1.46 $\AA$ and 1.64 $\AA$ respectively.}
\label{fig:interplanardistSI}
\end{figure*}

\begin{figure*}[hbt!]
\centering
\includegraphics[width=0.8\textwidth]{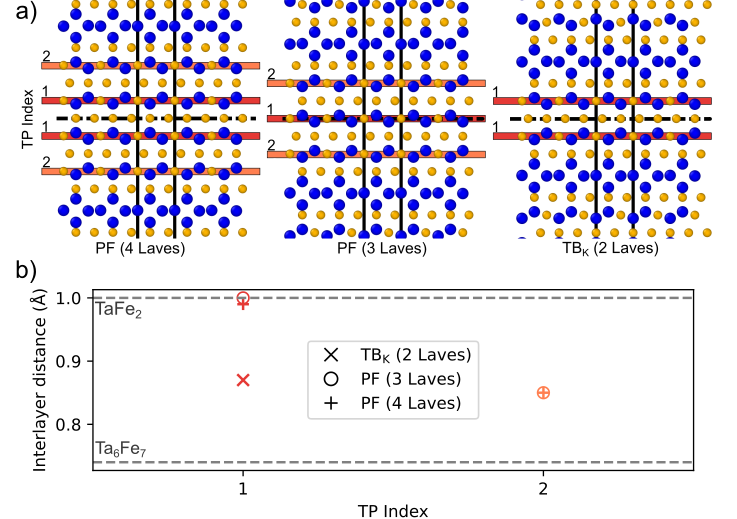}
\caption{DFT-simulated Triple-layer (TP) Ta-Ta interplanar distances of the Laves phase layers in PF(4 Laves), PF(3 Laves) and TB$_K$(2 Laves) of the  Ta$_6$Fe$_7$ $\mu$-phase, where each distinct TP layers are indexed from 1 to 2 depending on its distance from the mirror plane. For comparison, the TP Ta-Ta distances of TaFe$_2$ and Ta$_6$Fe$_7$ are shown as labeled dashed lines at 1.0 $\AA$ and 0.74 $\AA$ respectively.}
\label{fig:interplanardistSI2}
\end{figure*}

\end{document}